\documentclass{jaa}
\usepackage[usenames,dvipsnames]{color}
\usepackage{natbib}
\usepackage[hyperfootnotes=false]{hyperref}

\hypersetup{colorlinks=true, citecolor=blue, linkcolor=blue,  filecolor=blue, urlcolor=blue}
\usepackage[english]{babel}
\usepackage{epsfig}
\usepackage{epstopdf}
\usepackage{psfrag}
\usepackage{color}
\usepackage{graphicx}
\usepackage{fancyhdr}
\usepackage{graphics} 
\usepackage{soul}

\usepackage{amsmath}
\usepackage{upgreek}

\usepackage{xspace}
\usepackage{xcolor}

\usepackage{graphicx}

\newcommand{\Sint}{{S_{\rm int}}}
\newcommand{\Speak}{{S_{\rm peak}}}

\begin{document}\sloppy

\title{The spectral study of the faint radio sources in the ELAIS N1 field}


\author{Akriti Sinha\textsuperscript{*}, Sarvesh Mangla\textsuperscript{} and Abhirup Datta\textsuperscript{} }
\affilOne{\textsuperscript{}Indian Institute of Technology Indore, India\\}


\twocolumn[{

\maketitle

\corres{sinha.akriti44@gmail.com}


\begin{abstract}

Understanding the spectral properties of sources is crucial for the characterization of the radio source population. 
In this work, we have extensively studied the ELAIS N1 field using various low-frequency radio observations. For the first time, we present the 1250\,MHz observations of the field using the upgraded Giant Meterwave Radio Telescope (uGMRT) that reach a central off-source RMS noise of $\sim 12\,\mu$Jy\,beam$^{-1}$. A source catalogue of 1086 sources is compiled at $5\sigma$ threshold ($>60\,\mu$Jy) to derive the normalized differential source counts at this frequency that is consistent with existing observations and simulations. We present the spectral indices derived in two ways: two-point spectral indices and by fitting a power-law. The latter yielded a median $\alpha = -0.57\pm 0.14$, and we identified nine ultra-steep spectrum sources using these spectral indices. Further, using a radio colour diagram, we identify the three mega-hertz peaked spectrum (MPS) sources, while three other MPS sources are identified from the visual inspection of the spectra, the properties of which are discussed. In our study of the classified sources in the ELAIS N1 field, we present the relationship between $\alpha$ and $z$. We find no evidence of an inverse correlation between these two quantities and suggest that the nature of the radio spectrum remains independent of the large-scale properties of the galaxies that vary with redshifts.

\end{abstract}


}]


\doinum{12.3456/s78910-011-012-3}
\artcitid{\#\#\#\#}
\volnum{000}
\year{0000}
\pgrange{1--}
\setcounter{page}{1}
\lp{1}

\section{Introduction}

The investigation of the deep radio sky is crucial for examining the population of galaxies at different redshifts. This population is primarily composed of various sources, including star-forming galaxies (SFGs) and active galactic nuclei (AGN). 
Many galaxies are believed to have supermassive black holes at their centres, which power AGN. The relativistic jets from these AGN also provide feedback that shapes and affects the galaxy and the intergalactic medium. Therefore, we must comprehend AGN evolution in order to fully understand galaxies and their evolution \citep{Fabian2012}. In fact, the early evolutionary stages of a radio AGN are still debated \citep{Orienti2016,Bicknell2018}. Due to similarities between them at the kpc and pc scales, compact radio doubles discovered using Very Long Baseline Interferometry (VLBI) have been proposed as the ancestors of Fanaroff-Riley (FR) type I and type II radio-loud AGN \citep[RL\,AGN;][]{ODea1997}.

The differential source counts are utilised to investigate the nature of the extra-galactic sources and galaxy evolution \citep{Padovani_2011,Padovani_2015,Prandoni_2018}. 
It is observed that the deep radio sky is dominated by RL\,AGN population at high flux density regimes, whereas the population of SFGs becomes more dominant at lower flux density ends \citep[see][for e.g.]{Smolcic_2008,Padovani_2016, Smolcic_2017, Best_2023} as is reflected from the flattening of the normalised source counts. 
It also has been found that the fainter flux density regimes also comprise another source population called radio-quiet AGN (RQ\,AGN) \citep{Padovani_2009,Bonzini_2013} whose radio emission mechanism is still debated. Some studies like \cite{Miller_1993} suggested that these sources are mini-scaled versions of RL\,AGN while \cite{Sopp_1991} proposed the radio emissions from these sources are from the star-forming regions of the host galaxies. The recent study by \cite{Pannesa_2019} discuss a wide range of mechanisms responsible for radio emission in the RQ\,AGN population: AGN-driven wind, star-formation, jets with low power, coronal activity in the innermost accretion disk and free-free emission from the photoionised gas \citep[see][and references therein]{Pannesa_2019}. 

The low-frequency observations are essential for the detection of faint radio sources, also ultra-steep spectrum  (USS) sources, which are generally the radio galaxies at high-$z$ \citep{Best_2003,Miley_2008}. Complementary radio observations at low- and high-frequencies are crucial for the characterization of sources depending on their spectra \citep{Coppejans_2015,Mahony_2016_GPS}. This may lead to the identification of diverse source populations with different spectral properties. For example, USS sources \citep{Roettgering_1994}, gigahertz-peaked spectrum sources (GPS) \citep{Athreya_1999,ODea_1998}, and core-dominated RQ\,AGN \citep{Blundell_2007}.

Furthermore, a family of radio AGN called GPS, compact steep spectrum (CSS), and high-frequency peaked (HFP) sources have been proposed as the young counterparts of enormous RL\,AGN \citep{Hardcastle2019}. HFP sources have spectral peaks above 5 GHz with pc-scale linear sizes, GPS sources peak around 1 GHz with linear sizes $\leq 1$\,kpc, while CSS sources peak at low frequencies ($\leq$\,500 MHz) and have linear sizes of $\sim 1-20$\,kpc. Besides these classes of sources, there are recent studies that have found megahertz-peaked spectrum (MPS) sources with a similar spectrum that peak at frequencies below 1\,GHz \citep{Coppejans_2015, Callingham_2017}. As a result of cosmic evolution,  MPS sources are considered as the amalgamation of nearby CSS sources and GPS and HFP sources at high redshift whose turnover frequencies have been redshifted to lower frequencies below 1 GHz \citep{Coppejans_2016}.

\begin{table*}
    \centering
    \caption{Observation summary of the calibrator sources and target field (ELAIS N1) for three observing sessions}	
    \begin{tabular}[width=\columnwidth]{lr}
    \hline
    \hline
    Project code & 31\_072 \\
    Observation date & 26 Feb 2017 \\
                     & 27 \& 28 Mar 2017 \\
    \hline
    Bandwidth & 400 MHz \\
    Frequency range & 1.05-1.45 GHz \\
    Channels & 2048 \\
    Integration time & 16.1s \\
    Correlations & RR RL LR LL \\
    Flux calibrator & 3C\,286 and 3C\,48 \\
    Phase calibrator & 1634+627 \\
    Total on-source time &  $\sim$15\,hr (ELAIS N1)\\
    \hline
    Pointing centres & $13^{h}31^{m}08^{s}$  $+30^{d}30^{m}33^{s}$ (3C\,286) \\
                     & $16^{h}34^{m}34^{s}$  $+62^{d}45^{m}36^{s}$ (J1634+627) \\ 
                     & $16^{h}10^{m}00^{s}$  $+54^{d}40^{m}00^{s}$ (target) \\
                     & $16^{h}10^{m}00^{s}$  $+54^{d}22^{m}00^{s}$ (target) \\
                     & $16^{h}08^{m}13^{s}$  $+54^{d}31^{m}00^{s}$ (target) \\
                     & $16^{h}11^{m}47^{s}$  $+54^{d}31^{m}00^{s}$ (target) \\
                     & $16^{h}08^{m}12^{s}$  $+54^{d}48^{m}50^{s}$ (target) \\
                     & $16^{h}11^{m}48^{s}$  $+54^{d}48^{m}50^{s}$ (target) \\
                     & $16^{h}10^{m}00^{s}$  $+54^{d}57^{m}50^{s}$ (target) \\
                     & $01^{h}37^{m}41^{s}$  $+33^{d}09^{m}35^{s}$ (3C\,48) \\
    \hline                 
    \hline                 
    \end{tabular}
    \label{tab:obs}	    
\end{table*}

There are two debated scenarios for these sources: ``youth'' hypotheses, where these sources represent the young precursors of RL AGN \citep{Wilkinson_1994,ODea_1998,Murgia_1999,Orienti_2006,An_2012} or the ``frustration'' model, where the compact sizes are caused by a dense medium surrounding the nucleus \citep{VanBreugel_1984,Bicknell_1997}.
By determining whether synchrotron self-absorption (SSA) or free-free absorption (FFA) is in charge of the change in the radio spectrum, one can determine whether a GPS, CSS, or HFP source is young or frustrated. A source will typically have optical thickening at low frequencies due to SSA, which is caused by the relativistic electrons themselves and has a characteristic spectral index limit of 2.5 ($S \propto \nu^{\alpha}$) below the spectral turnover. Moreover, it has been found that FFA presumably dominates the source's spectrum below the turnover within small spatial scales surrounded by the dense circum-nuclear medium. Most recently,  \cite{Keim2019} have studied 6 sources selected from \cite{Callingham_2017} in the Galactic and Extragalactic All-sky Murchison Widefield Array survey and suggest FFA as to be the plausible absorption mechanism for their sources.

In this paper, we analyse the uGMRT data at 190\,MHz and 1.2\,GHz of the ELAIS N1 field to derive the spectral indices of the sources in the 400\,MHz radio catalogue (see \cite{Arnab2019}) by performing the spectral energy distribution (SED) fitting. Furthermore, for the first time, we identify the MPS sources in the field using a radio colour diagram and spectrum. We present some of their properties here. 
Using the measured spectral indices from the SED fitting, we also identify USS sources in the field and discuss the variation of $\alpha$ as a function of $z$ for the sources in the field.

\section{Observations and Analysis}

The focus of this section is on the radio observations carried out in the ELAIS N1 field, which have been utilized in our study. We have utilised the uGMRT data at 400\,MHz from \cite{Arnab2019} and \cite{Akriti_2022} and extended the analysis on investigating the spectral properties of the faint radio sources in the field. Here, we present the 1250\,MHz uGMRT observations of the ELAIS\,N1 field followed by discussing other radio continuum data that have been used.

\subsection{uGMRT L-Band Data}

For this study, we used archival observations of the ELAIS N1 field ($\alpha_{2000} = 16^{h}10^{m}00^{s},\, \delta_{2000} = 54^{d}36^{m}00^{s}$) taken using uGMRT during GTAC cycle 31 (Project Code: 31\_072). These observations are taken for three sessions and the total time for observation (for seven pointings) was 20\,hr (including calibrators), which centred around 1250\,MHz with a total bandwidth of 400\,MHz. The ELAIS N1 field was observed during the night time for all three days on 26\,Feb and 27-28\,March 2017. Either 3C\,286 or 3C\,48 or both are observed at the beginning and at the end for each observing day. The phase calibrator 1634+627 near the target field is observed for five minutes in between 3 or 4 pointings of the ELAIS N1 field. The total on-source time for every pointing is nearly 130 min. We summarize the observation details in Table \ref{tab:obs}. In the following sub-sections, we describe the data reduction and imaging procedure for creating a mosaic image of the ELAIS N1 field.

\begin{figure*}[ht]
    \centering
    \includegraphics[width=1.9\columnwidth]{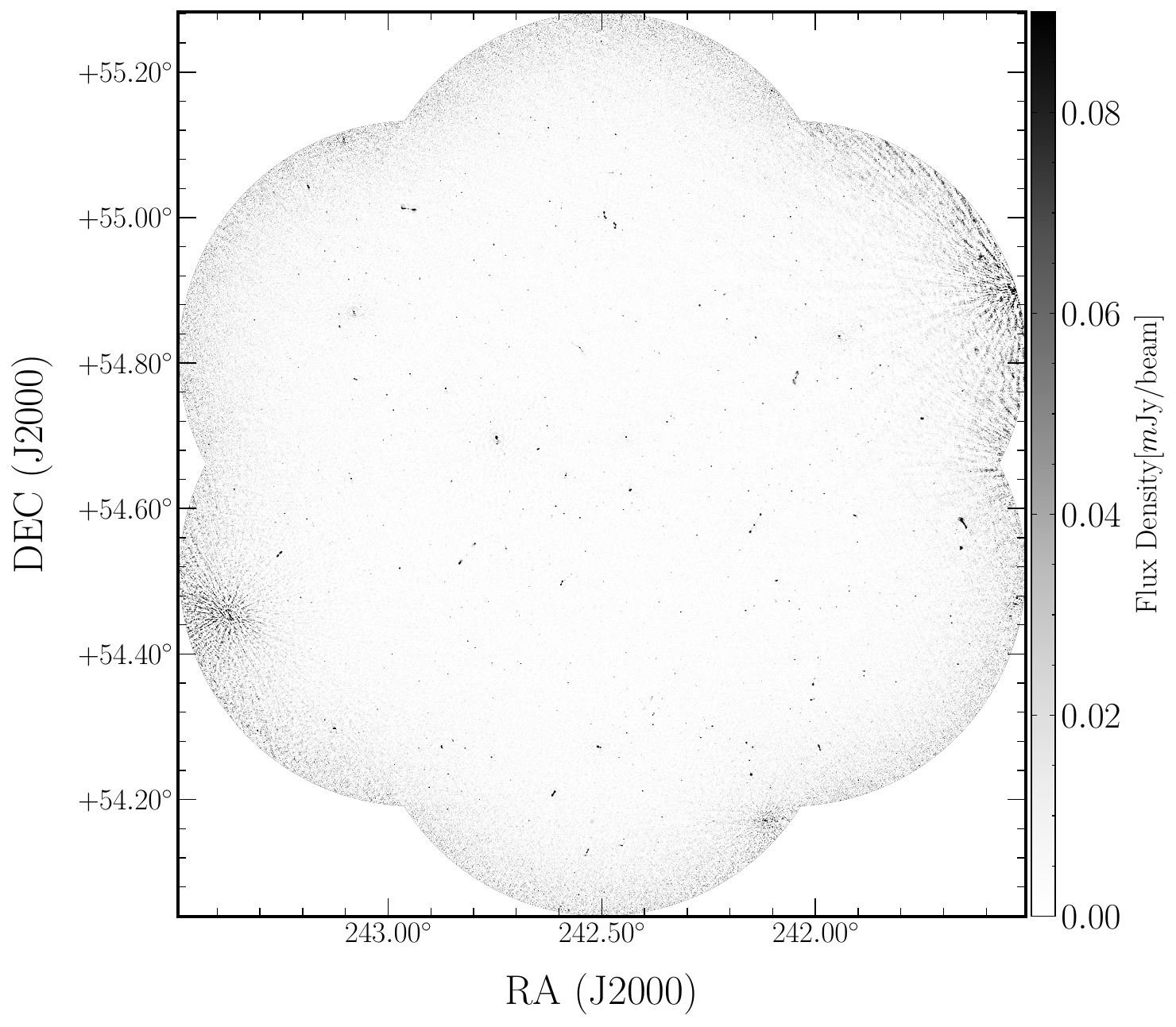}
    \caption{The uGMRT mosaicked map of the ELAIS\,N1 field at 1.25\,GHz with the central off-source RMS of 12\,$\upmu$Jy\,beam$^{-1}$. The resolution of the image is $2''.3\times1''.9$.}
    \label{fig:image1}
\end{figure*}

\subsection{Data reduction}

For the pre-processing of the data, like flagging, radio frequency interference mitigation and calibration, we used ACAL\footnote{\href{https://github.com/Arnab-half-blood-prince/uGMRT\_Calibration\_pipeline}{https://github.com/Arnab-half-blood-prince/uGMRT\_Calibration\_pipeline}} which is a \texttt{CASA}\footnote{Common Astronomy Software Applications (\url{https://casa.nrao.edu/})} based pipeline. This pipeline only perform direction-independent (DI) calibration solutions and at 1.25\,GHz ionosphere does not corrupt solutions that much even at low latitude. Each data went through the pipeline for calibration and finally all the calibrated data of each pointing (see Table \ref{tab:obs}) is used to get the complete image. A brief overview of the pipeline procedure will be explained in \textcolor{blue}{Sinha \& Datta (under review)}. Now, we split the calibrated data of all seven target pointings individually for imaging and self-calibration.

\subsection{Imaging \& Self-calibration}
The calibrated data for all the seven pointings is used for imaging individually using \texttt{WSCLEAN} \citep{Offringa_2014}. 
Multi-scale wide-bandwidth deconvolution, as described by \cite{Offringa_Smirnov_2017} is used to capture the variation in sky brightness across spatial scales. In this study, we chose Briggs robust parameter value of $-1$, which ensures uniform weighting across the data. This choice produces a central Gaussian point spread function (PSF), which minimizes the presence of broad wings. To incorporate any bright sources that are located outside the field of view, we generated a large sky map with a size of approximately $1.1\deg^2$ with each pixel having the size of $0.5''$. In order to optimize the image quality, we perform an initial round of imaging using the auto-masking algorithm of \texttt{WSCLEAN} with 50k iterations, down to a significance level of 7$\,\sigma$. The Multi-Frequency Synthesis (MFS) image is employed to generate a mask for a subsequent round of imaging. This is a standard procedure to fill the \texttt{MODEL\_DATA} column, which is necessary for self-calibration because a model column will be less susceptible to imaging artefacts, resulting in better solutions for subsequent self-calibration loops.

By utilising the above-explained method, we performed four rounds of phase-only self-calibration procedure with solution intervals (solint) as 8\,mins, 6\,mins, 4\,mins and 2\,mins. After incorporating the latest self-calibration solutions, the final image of the target field is generated. However, we did not perform amplitude-only or amplitude-phase self-calibration any further.\par
After performing self-calibration and separate imaging for each pointing, we applied correction for the frequency-dependent uGMRT primary beam model\footnote{\url{http://www.ncra.tifr.res.in/ncra/gmrt/gmrt-users/observing-help/ugmrt-primary-beam-shape}}. In this process, we utilized 20$\%$ of the primary beam response to correct the image for each pointing.     Subsequently, we used \texttt{MONTAGE}\footnote{\url{http://montage.ipac.caltech.edu/}}, to create a linear mosaic of the seven pointings, resulting in the one final combined image. In this process, each primary-beam corrected image is weighted by the square of the primary-beam pattern, which is considered proportional to the noise variance image. The overall mosaic of the ELAIS N1 field, covering an area of approximately $\sim2.4\,\deg^2$, is depicted in Figure \ref{fig:image1}. Our analysis achieved a minimum central off-source RMS noise of $\rm12,\upmu Jy,beam^{-1}$ with a beam size of $2''.3\times1''.9$.

\begin{table*}
\caption{Sample of the source catalogue at 1250\,MHz from the uGMRT observations of the ELAIS N1 field.} \label{tab:cat}
\scalebox{1.0}{
\begin{tabular}{l l l l l l l l l l}
\hline \hline
Id & RA  & DEC & Total\_flux &  Peak\_flux & Major & Minor & PA & RMS  \\

() & (deg)  & (deg)  & (mJy)& (mJy $\mathrm{beam}^{-1}$) & (arcsec) & (arcsec) & (degree) & (mJy $\mathrm{beam}^{-1}$) \\
(1) & (2)  & (3)    & (4) & (5)     & (6) & (7) & (8) & (9) & \\
 \hline
\hline
  0 & 243.4352 & 54.6886 & 1.944 & 1.451 & 2.5 & 2.3 & 136.64 & 0.085\\
  1 & 243.4006 & 54.5029 & 1.099 & 0.679 & 2.9 & 2.4 & 92.45 & 0.06\\
  2 & 243.3881 & 54.4056 & 0.544 & 0.561 & 2.4 & 1.7 & 80.06 & 0.071\\
  3 & 243.3811 & 54.4551 & 78.51 & 44.392 & 2.7 & 2.6 & 42.41 & 0.408\\
  4 & 243.3793 & 54.4546 & 36.101 & 25.306 & 2.9 & 2.1 & 79.86 & 0.408\\

\hline
\end{tabular}}
\begin{flushleft}
$^\dagger$The electronic version of the catalogue is available where the columns include the source ids, positions, flux densities and peak flux densities along with their respective errors. It also includes the sizes, position angle and the local RMS noise.
\end{flushleft}
\end{table*}

\begin{figure}
    \centering
    \includegraphics[width=8cm, height = 7cm]{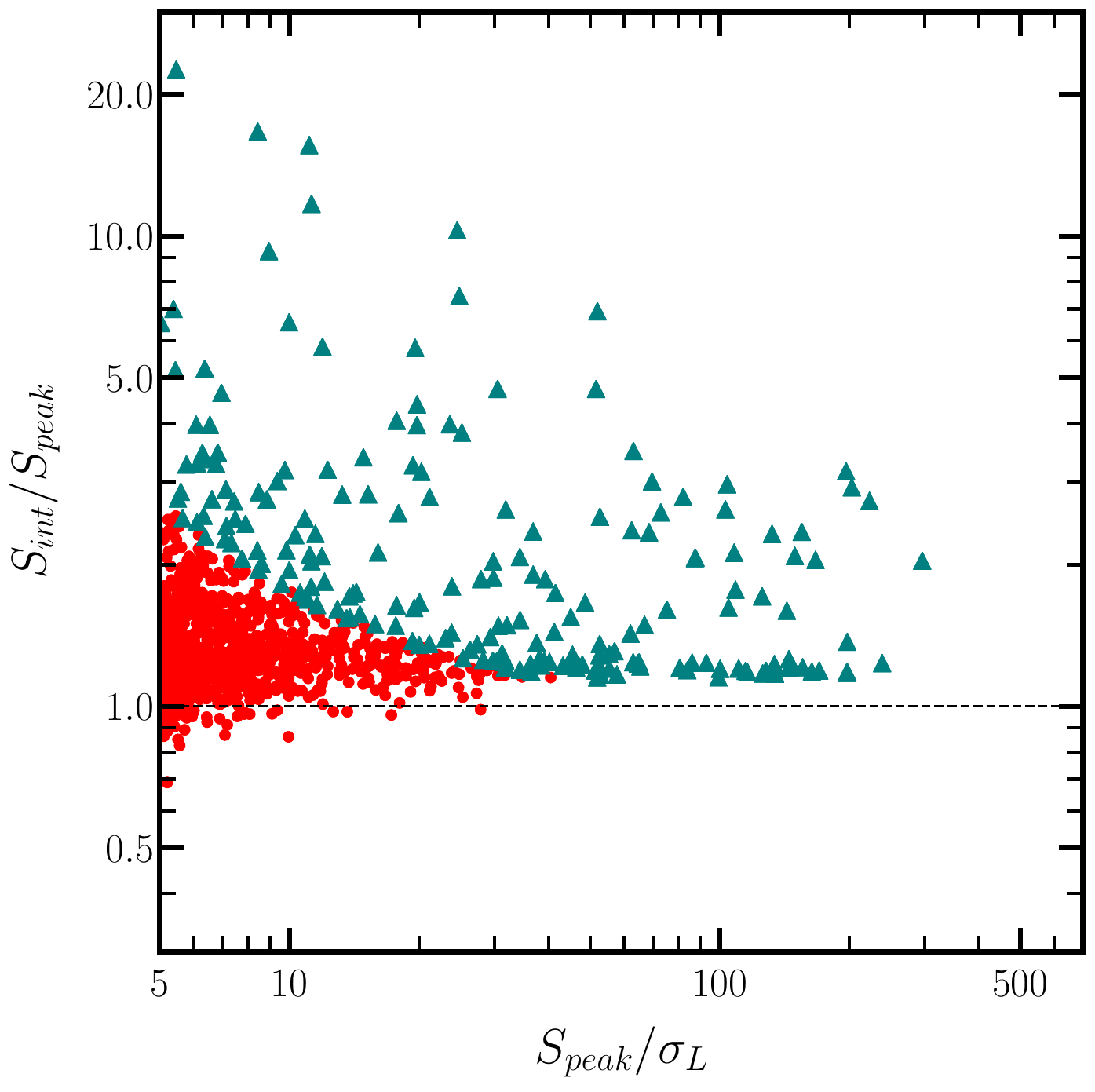}
    \caption{The variation of the ratio of the integrated flux to the peak flux with the SNR of the source. The red and teal-coloured sources represent the point and resolved sources, respectively. }
    \label{fig:resolve}
\end{figure}

\subsection{Source Catalogue}

In this study, we have used P{\tiny Y}BDSF \citep{Mohan_Raffrey2015} software to create a catalogue of sources and characterize them. We employed a sliding box window, \texttt{rms\_box} = (170,40), on the final mosaicked image. To avoid counting bright artefacts as real sources, a smaller box of size (38,8) was used around them. These bright regions are selected using an adaptive threshold of 150$\sigma_{\rm RMS}$. 
P{\tiny Y}BDSF was then used to identify continuous regions of emission above a pixel threshold (\texttt{thresh\_pix} = $5\sigma$ and \texttt{thresh\_isl} = $3\sigma$) and model each region by fitting multiple Gaussian components. 

Nearby Gaussians on the same island are grouped together to form a single source using P{\tiny Y}BDSF. Thus, total flux is measured as the sum of all the Gaussian fluxes, while the uncertainty is calculated by adding individual Gaussian uncertainties in quadrature. The position of the source is given by the centroid of the source. 
The PSF of the image may vary from its original value of the restoring beam. This is taken care of by using the parameter \texttt{psf\_vary\_do = True}.

Using P{\tiny Y}BDSF, we generated a source catalogue consisting of a total of 1086 sources above $> 60\,\upmu$Jy flux density ($5\sigma$). 
In Table \ref{tab:cat}, we list a sample of the source catalogue while the complete catalogue will be available with the electronic version of the paper. We have compared our catalogue with other radio catalogues to investigate the positional and flux accuracies. This is discussed in \ref{sec:flux_position}
We will discuss the point and resolve source identification in the following subsection.

\subsection{Classification of sources}

Because of various effects like time and bandwidth smearing, a source may get elongated in the image plane. Thus, making it difficult to identify the actual point and extended sources. For exceptional noise-free cases, the ratio of the integrated to the peak flux densities, $S_{\rm int}/S_{\rm peak}> 1$ can be used to identify resolved sources. Figure \ref{fig:resolve} presents the variation of $\Sint/\Speak$ with $S_{\rm peak}/\sigma_{\rm l}$, where $\sigma_{\rm l}$ represents the local RMS here. It is deemed from the Figure that the distribution is skewed at low SNR, which could be possible because of the noise variation and calibration uncertainties.

Following \cite{Franzen_2015,Franzen_2019}, we identify the resolved and point-like source in our sample. The RMS was estimated as:
\begin{equation}
    \sigma_{\rm R} = \sqrt{ \Big(\frac{\sigma_S}{S_{\rm int}} \Big)^2 + \Big(\frac{\sigma_{S_{\rm peak}}}{S_{\rm peak}}\Big)^2}
\end{equation}

Thus, a source is classified as resolved if $ln(S/S_{\rm peak}) > 3\sigma_{\rm R}$ \citep{Franzen_2015}. In this way, we identify 204 as extended and 882 sources as point-like from our 1250\,MHz uGMRT catalogue.

\section{Source Counts at 1.25\,GHz}

It is essential to understand the population distribution as a function of flux density, especially at low radio frequencies. Further, it has been found from both observations \citep{Padovani_2015,Smolcic_2017,Best_2023} and simulations \citep{Bonaldi_2019} that the population of SFGs and RQ\,AGN dominate at faint fluxes . We have derived the normalised differential source counts for our uGMRT catalogue at 1.25\,GHz down to 60\,$\mu$Jy. However, direct measurement of source counts may have biases based on false detection rates, incompleteness, Eddington bias etc. A few of these correction factors used are discussed below in brief \citep[following][]{Williams_2016}.  

\begin{figure}
    \centering
    \includegraphics[width = 7.5cm]{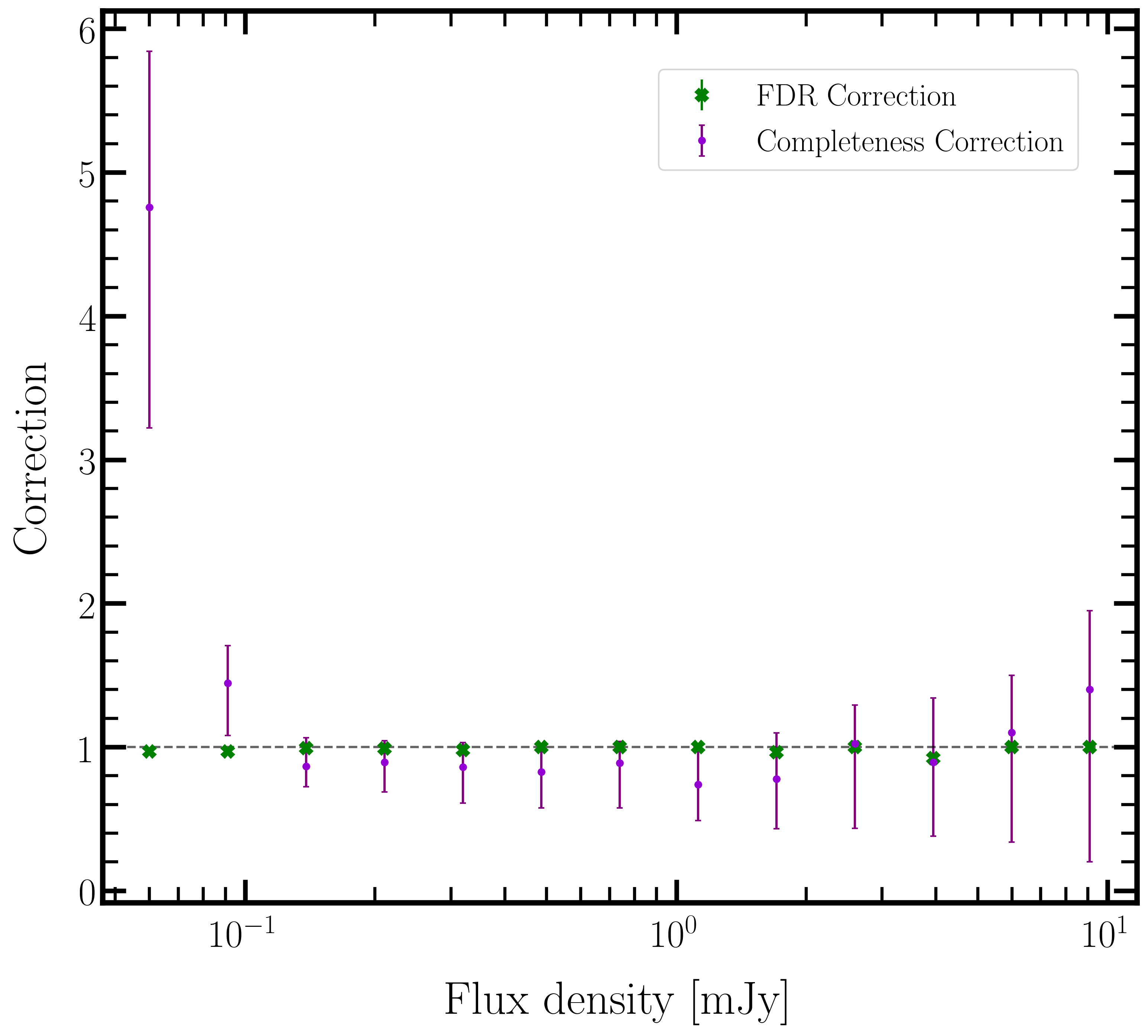}
    \caption{Correction factors are shown in green and purple because of FDR and completeness, respectively.}
    \label{fig:corrections}
\end{figure}

\subsection{FDR}

There could be possible false detection of sources due to noise and artefacts in the \texttt{PyBDSF} compiled catalogue. The total count of spurious detections is referred to as false detection rates. Assuming a symmetrical distribution around zero, \texttt{PyBDSF} detected spurious sources and negative sources (from inverted image) will be equal.
This can be quantified by running \texttt{PyBDSF} with the same parameters as applied for the original image. 
There were  20 sources with negative peaks less than the $-5\sigma$ threshold. Following the method described in \cite{Hale2019}, we measured the correction for each flux bin and multiplied the corresponding source count.

\subsection{Completeness}

The \texttt{PyBDSF} source catalogue is not entirely complete due to factors that can lead to both overestimation and underestimation of source counts. Incompleteness refers to the inability to detect sources above a given flux density limit due to varying noise in the image. Eddington bias redistributes low flux density sources into higher fluxes, resulting in a boost in source counts in the faintest bins. Resolution bias reduces the detection probability of extended sources, leading to a reduction in source counts. The completeness of the catalogue was quantified by injecting 800 sources whose fluxes were derived as dN/dS $\propto S^{-1.6}$ \citep{Intema2011, Williams2013}, into the primary beam-corrected image and extracting sources using PYBDSF. We have simulated these sources such that 650 are point sources, i.e., their major and minor axes lie in the range $1.3''-2''$, while the remaining 150 are extended sources, i.e., major and minor axes are in the range $2''-15''$. The sources are selected from a uniform distribution within these ranges. The ratio of the number of points and extended sources is taken nearly similar to that of the original 1.25 GHz catalogue. Following the general trend of previous observations \citep{Williams_2016,Chakraborty_2020,Mandal_2021}, we are within the uncertainty limits. This simulation was performed 100 times for robust estimation of the completeness factor. The completeness correction factor was calculated as the ratio of the number of injected sources with the number of recovered sources \citep{Hale2019} for each flux bin, which accounts for both resolution bias and Eddington bias. Finally, the median values of the completeness factor from the 100 simulations in each flux bin are used to measure the completeness correction factor. Fig. \ref{fig:corrections} shows the variation of correction factors due to FDR and completeness with the flux bins in green and violet, respectively.

\begin{figure}
    \centering
    \includegraphics[width = \columnwidth]{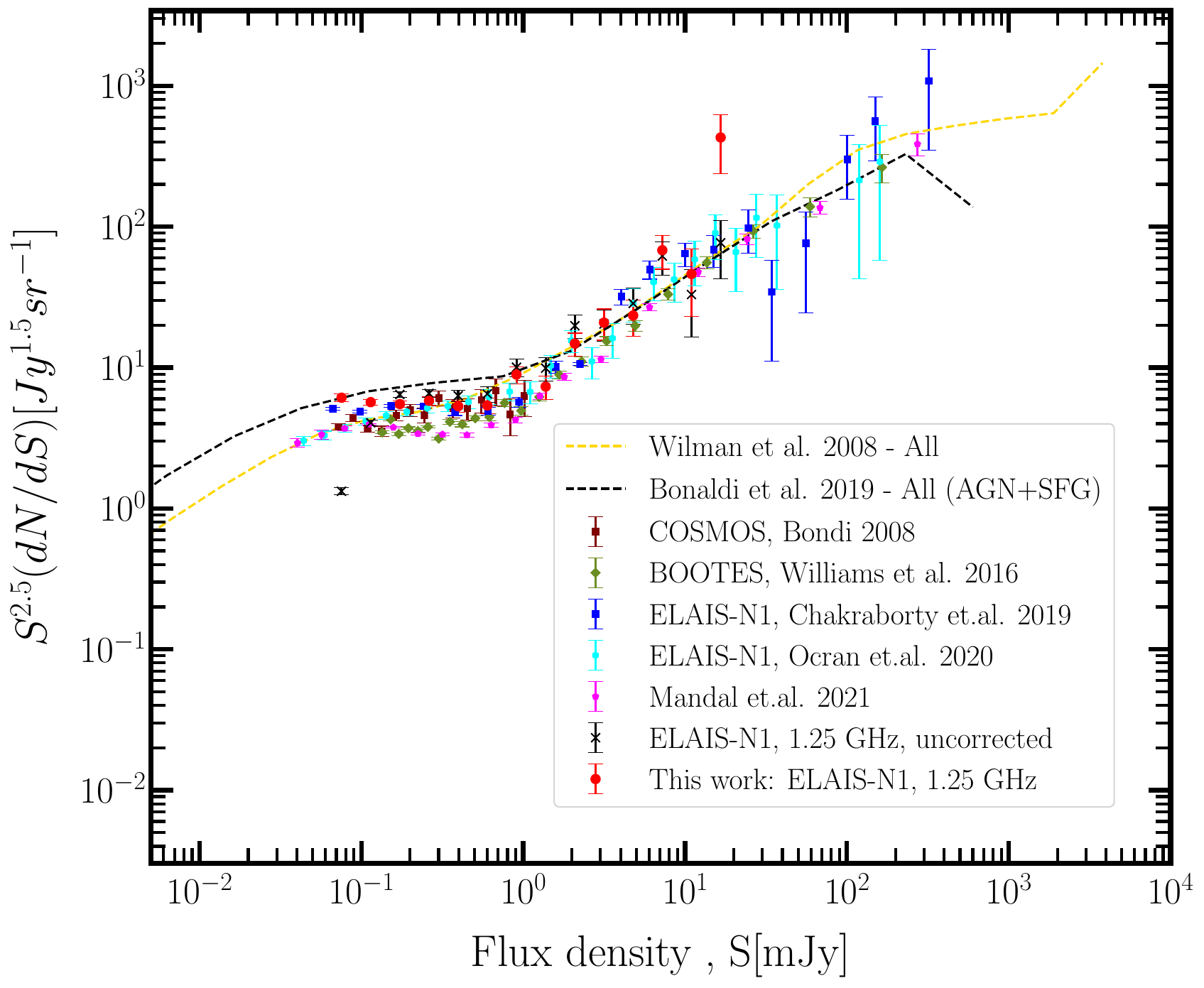}
    \caption{Euclidean Differential source counts for the sources in the ELAIS N1 field using 1250\,MHz uGMRT observations. The red points represent the corrected source counts here. The black crosses are the uncorrected source counts and are shown here only for comparison.  }
    \label{fig:dnds}
\end{figure}

\subsection{Differential Source Count}

We obtained normalized differential source counts at 1250\,MHz from the uGMRT catalogue and adjusted for false detection rates and completeness using correction factors for each flux bin. Fig. \ref{fig:dnds} shows the corrected, normalised source counts in red points and the uncorrected counts in black crosses. We employed 11 logarithmically-spaced bins according to the flux density up to 60\,$\upmu$Jy ($5\sigma$). The errors estimated for the counts in each bin are Poisson errors. 

In Fig. \ref{fig:dnds}, we also show the source counts from various observations and simulations in the literature. All these values are scaled to 1250\,MHz with $\alpha=-0.8$. Here, we have used the source counts for the same field, ELAIS N1 from \cite{Arnab2019} at 400\,MHz and \cite{Ocran_2020} at 610\,MHz. The recent work by \cite{Mandal_2021} is also shown who have combined the LOFAR catalogue at 150\,MHz from the three fields, the ELAIS N1, the Bo\"{o}tes and the Lockman Hole field. For comparison, we also have used the differential source counts from \cite{Bondi2008} for the COSMOS field and from \cite{Prandoni_2018} for the Lockman Hole field, both at 1.4\,GHz.
Further, we used the $S^3-$SKADS simulations  \citep{Wilman_2008} and T-RECS simulations \citep{Bonaldi_2018}, who have measured the source counts at 1.4\,GHz.

Our measures of the differential source count at 1.25\,GHz are consistent with the literature and flatten roughly around 1\,mJy. At lower flux densities, it is observed that the counts at 1.25\,GHz have a slight positive offset when compared to the results of \cite{Williams_2016} or \cite{Mandal_2021}. However, these are well within the boundary of the two simulations from SKADS and T-RECS. 
The mentioned simulations suggested an increase in the population of faint radio sources like SFGs and RQ\,AGN is reflected in the flattening of source counts below 1\,mJy. Also, the recent observations from Sinha \& Datta (under review) have confirmed the dominance of these source populations for the Bootes field using 400\,MHz uGMRT observations.
When compared  at higher flux densities, the measured counts are consistent with sky models and observations.

\begin{figure}
    \centering
    \includegraphics[width = \columnwidth]{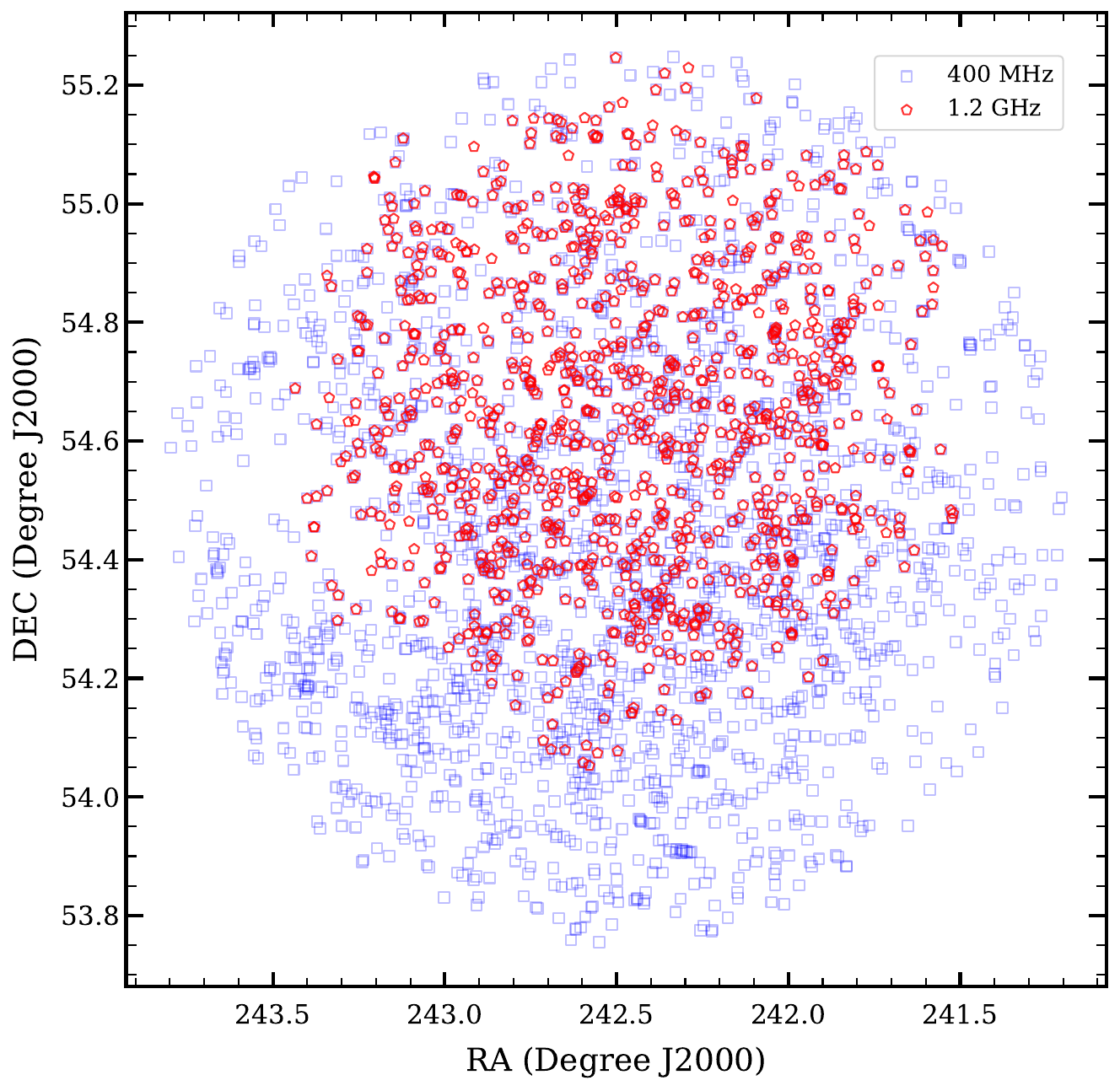}
    \caption{The source distribution in the ELAIS\,N1 field at 1.2\,GHz in red and 400\,MHz in blue using uGMRT.  }
    \label{fig:umgrt}
\end{figure}

\begin{table*}
    \centering
    \caption{Details of the catalogues considered. The columns represent the catalogue, frequency in MHz, resolution, corresponding RMS noise and their $5\sigma$ sensitivity in mJy. The last two columns represent the total number of sources in a survey that have a counterpart in the 400\,MHz uGMRT catalogue and their corresponding percentages, respectively. } 
    \begin{tabular}{ccccccc}
     \hline \hline
    Catalogue & Frequency & Resolution & $\sigma$  & $S_{\rm limit}$  & Size & Percentage \\
        & (MHz) & (arcsec) & (mJy) & (mJy) & &  \\
    \hline
    uGMRT & 400 & $4''.5$ & 0.015 & 0.075 & 2528$\dagger$ & 100 \\
    uGMRT & 1250 & $2''.0$ & 0.012 & 0.06 & 853 & 34 \\
    LOFAR & 150 & $6''.0$ & 0.02 & 0.10 & 2225 & 88 \\
    GMRT & 612  & $6''.0$ & 0.008 & 0.04 & 1518 & 60 \\
    \hline
    \end{tabular}
    \begin{flushleft}
        \footnotesize{\textit{Note:} The values in column four are the RMS noises as measured in the central region of the image. $\dagger$ All other catalogues are matched to these 2528 sources that are compiled above $6\sigma$ threshold (see \ref{sec:survey}).}
    \end{flushleft}
    
    \label{tab:survey}
\end{table*}

\section{Other Radio Continuum Data}\label{sec:survey}

Our study has relied upon the source catalogues derived from radio observations of the ELAIS N1 field at three distinct frequencies further: 146\,MHz obtained from LOFAR \citep{Sabater_2021}, 400\,MHz obtained from uGMRT \citep{Arnab2019}, and 612\,MHz obtained from GMRT \citep{Chakraborty_2020}. In this work, we have used the 400\,MHz uGMRT catalogue as the base catalogue that contains 2528 sources above the $6\sigma$ threshold with point source sensitivity $\geq 100\upmu$Jy. The ELAIS\,N1 field was observed in the frequency range 300-500\,MHz and the final image reached an RMS noise of $15\,\upmu\rm{Jy\,beam^{-1}}$ covering a sky area of $1.8\,\deg^2$. \cite{Arnab2019} describes the observations, calibration procedures and catalogue generation in detail.
Besides, we have also utilised the in-band catalogues centred at frequencies 325, 375, 425, and 475 MHz from \cite{Akriti_2022}, wherever available. 
We direct the reader to the specified references for a comprehensive account of the data analysis process and the catalogue creation. These sources provide a detailed description of the methodology and procedures utilized in this study, offering a more in-depth understanding of the data analysis and catalogue generation.
We have used a search radius of 3 arcsecs to determine the counterpart of sources in other catalogues except for 1250\,MHz where we have used a search radius of 2 arcsecs. In Figure \ref{fig:umgrt}, we show the overlay of the uGMRT 1250\,MHz catalogue in red points with the 400\,MHz catalogue in blue points. Table \ref{tab:survey} lists the salient features of the different surveys used in this work, with their corresponding total number of 400\,MHz uGMRT counterparts.

\subsection{Redshifts and AGN/SFG Classification }

We have used the redshift information from BOSS spectroscopy \citep{Bolton_2012}, the LOFAR photometric redshifts \citep{Duncan_2021} and the SWIRE redshift catalogue \citep{Rowan-Robinson2013} for our purpose \citep[see][for details]{Akriti_2022}. In this way, we have redshift information for 2319 sources in the 400\,MHz uGMRT catalogue. Moreover, we have used the classified sources as AGN (23.9\,\%) and SFGs (76\,\%) from \cite{Akriti_2022} based on the BOSS spectroscopy, IRAC colour classification, radio luminosity and observed $q$ values, i.e., logarithmic of the ratio of luminosities in the infrared and radio wavebands. It should be noted here that the sources in 400\,MHz catalogue with redshift measurements are only considered for AGN/SFGs identification. All sources are classified as SFGs that are not identified as AGN from the mentioned classification schemes.

\begin{figure*}
 \centering
 \includegraphics[width=8.5cm]{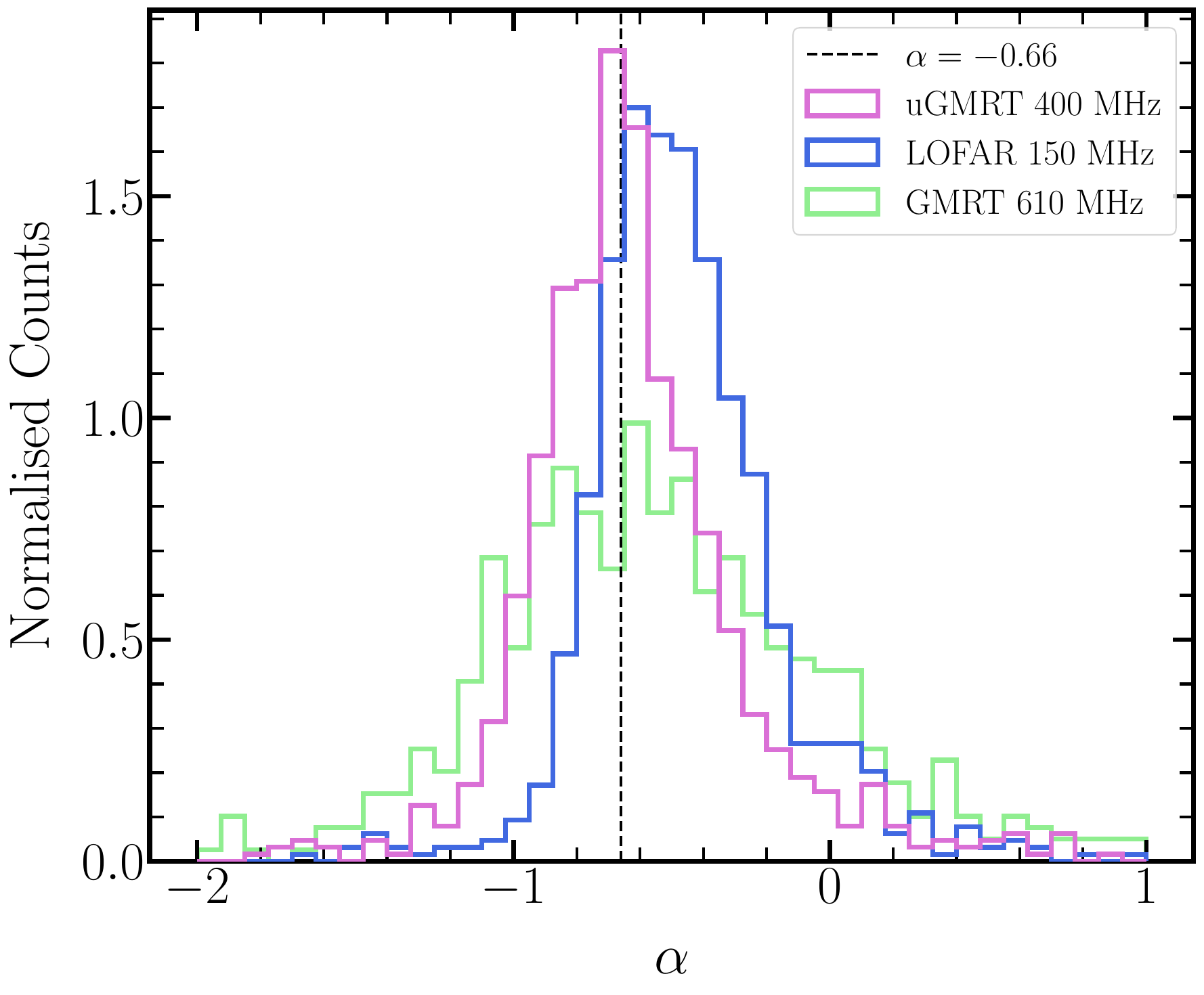}
 \includegraphics[width=8.5cm]{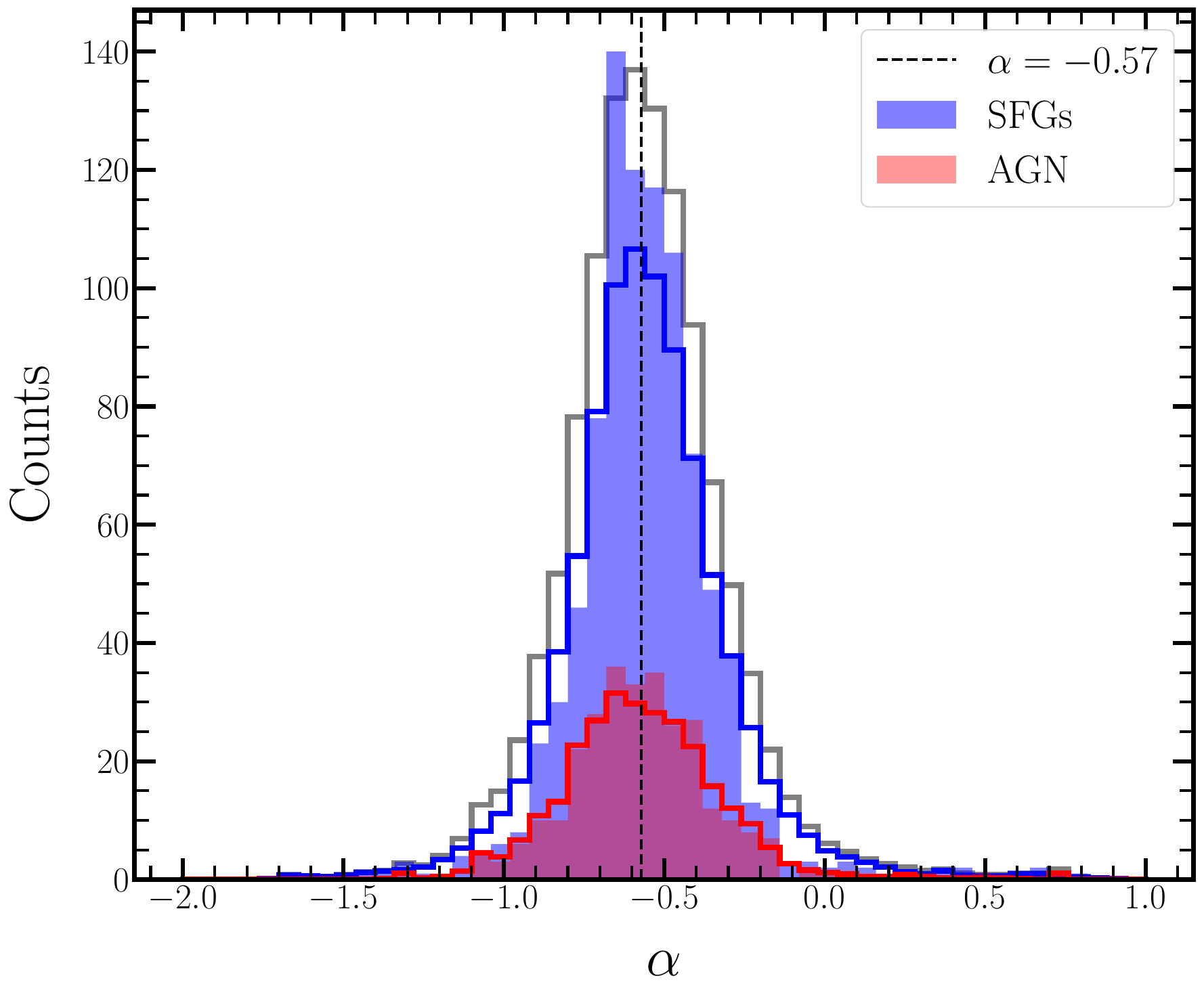}

 \caption{Spectral index distribution (see section \ref{sec:alpha}). \textit{Left:} Obtained by measuring two-point spectral indices between 1250\,MHz with 400, 146 and 612\, MHz. \textit{Right:} Derived by power-law fitting for each source. The dashed line indicates a median $\alpha$ value of -0.57. The filled histograms present the distribution of SFGs and AGN in blue and red, respectively. While the open histograms denote the distribution obtained by accounting for the errors in $\alpha$ values.  }
 \label{fig:alpha-hist}
\end{figure*}

\section{ Spectral Properties}\label{sec:spectral}

The spectral properties and radio SEDs offer valuable insights into the dominant emission mechanisms of different types of sources \citep{Prandoni_2010, Singh_2018}. Spectral indices help us to classify the radio source populations at higher wavelengths. This information can be crucial in identifying and classifying various astrophysical sources, such as AGN, SFGs, and radio relics, among others. A simple power-law model ($S\propto \nu^\alpha$) with negative $\alpha$ implies the dominance of radio synchrotron emission commonly observed in radio galaxies. Whereas spectra with positive $\alpha$ are indicative of very young compact sources, also called peaked spectrum sources. Interestingly, an $\alpha$ value of approximately 0 suggests two possibilities: optically thick synchrotron emission in core-dominated AGN or optically thin free-free emission in SFGs that dominates at high frequencies \citep[$\gtrsim$30\,MHz;][]{Condon_1992}. Therefore, by analyzing the shape and characteristics of SEDs, it becomes possible to gain insights into the underlying physical processes driving the observed emission.

We performed a detailed analysis of the spectral properties for sources in the ELAIS N1 by comparing their flux densities at various radio frequencies. 
We assumed synchrotron power-law distribution with a single spectral index unless specified. Thus, the spectral index ($\alpha$) were determined as:
\begin{equation}
    \alpha = \frac{\rm{log}(S_1/S_2)}{\rm{log}(\nu_1/\nu_2)}
    \label{eq:alpha}
\end{equation}
where, $S_1$ is the flux density at the frequency $\nu_1$ and $S_2$ is the flux density at $\nu_2$. The focus of this section is to analyze the spectral features of the sources and classify them based on their spectral properties.

\subsection{Spectral Index Distribution}\label{sec:alpha}
In order to analyze the spectral index distribution of the sources in the ELAIS N1 field, we calculated the two-point spectral indices (equation \ref{eq:alpha}) using 1.2\,GHz with 400\,MHz, 150\,MHz and 610\,MHz. The total number of sources that matched with our L-Band catalogue to estimate spectral index distribution are: 853 (400\,MHz uGMRT), 858 (150\,MHz LOFAR) and 548 (610\,MHz GMRT). 
Figure \ref{fig:alpha-hist} (left) shows the normalised histogram of the spectral indices estimated from the above-mentioned catalogues. 

The median spectral value for the frequency range 400\,MHz--1.2\,GHz is $-0.66\pm0.17$ and is shown with the black dashed line in the left panel of Figure \ref{fig:alpha-hist}. The median absolute deviation (MAD) was used to estimate the error on the median values for reference. We estimated a median value of $-0.50\pm0.16$ using 146\,MHz--1.2\,GHz and $-0.58\pm0.34$ using 612\,MHz--1.2\,GHz, respectively.

Furthermore, we extended a similar analysis as described in \cite{Akriti_2022} to measure the spectral indices by employing a power-law fit of the form $S\propto\nu^\alpha$. In brief, we divided the uGMRT data into four subbands between 300-500 MHz and ensured a minimum of three data points among all frequencies for robust SED fitting. Additionally, we conducted Monte Carlo simulations and drew 1000 random realizations for each frequency to account for flux density uncertainties. Each realization was then fitted to obtain 1000 $\alpha$ values, with the mean calculated as the $\alpha$ value for each source. Figure \ref{fig:alpha-hist} (right-panel) illustrates the $\alpha$ distribution for the sample of uGMRT sources in grey, with SFGs and AGN identified in \cite{Akriti_2022} shown in blue and red, respectively. The median $\alpha$ value for all sources was $-0.57\pm0.14$, with median values of $0.58\pm0.13$ and $-0.59\pm0.14$ for SFGs and AGN, respectively. These results are consistent with previous findings and support the trend of flattening spectral indices at lower frequencies, as reported in \cite{Coppejans_2015} and other relevant literature \citep{An_2023}.

\begin{figure*}
 \centering
 \includegraphics[width=1.5\columnwidth]{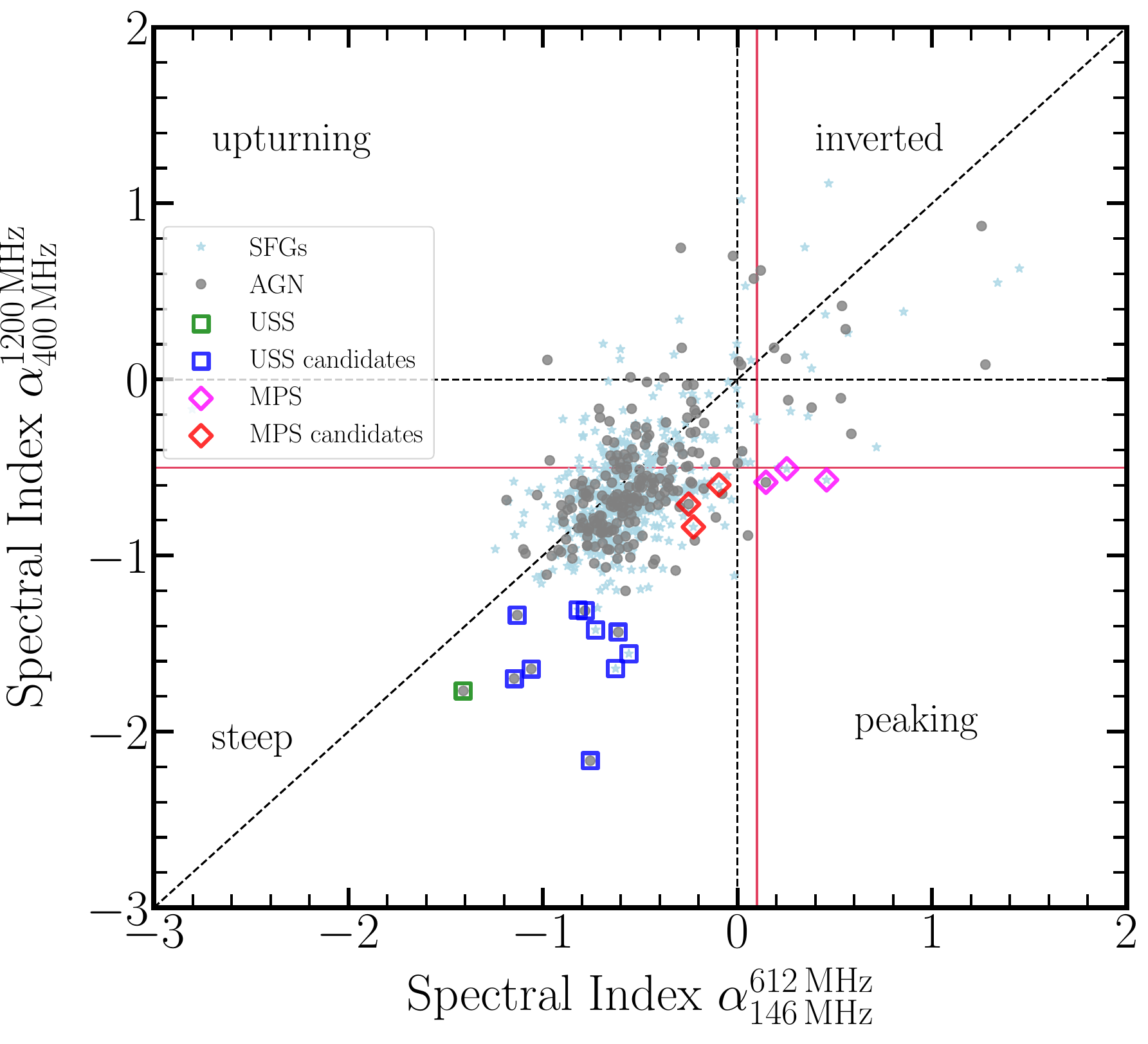}
 \caption{Radio colour-colour diagram where SFGs are represented by light-blue star symbols, and AGN are indicated by grey circle symbols. The figure is divided into four quadrants with dashed black lines that represent sources that may have upturning, inverted, steepening and peaking properties in their spectra (see section \ref{sec:spec_class}). The red lines represent spectral index limits used to identify the MPS sources in the field. The open squares and open diamonds represent USS and MPS sources, respectively (see text for details).}
 \label{fig:radio_color}
\end{figure*}

\begin{table*}
\centering
 \caption{Ultra Steep Spectrum Sources obtained by SED fitting. The ID column represents the source ID from the 400 MHz uGMRT catalogue. The spectral indices obtained by the power-law fit ($\alpha_{\rm fit}$) method and from the Monte Carlo method ($\alpha_{\rm MC}$) are also mentioned along with their redshifts in the last column.  } \label{USS}
\begin{tabular}{cccccc}
 \hline \hline \\
 ID & RA & DEC & $\alpha_{\rm fit}$ & $\alpha_{\rm MC}$ & $z$ \\ [1ex]
\hline  

50   & 243.6005 & 54.7216 & $-1.65 \pm 0.04$ & $-1.65 \pm 0.03$ & 0.34\\
1180 & 242.5347 & 54.3207 & $-1.30 \pm 0.13$ & $-1.32 \pm 0.13$ & 1.02\\
1335 & 242.4393 & 53.8292 & $-1.38 \pm 0.69$ & $-1.40 \pm 0.70$ & 0.63\\
1344 & 242.4347 & 53.9631 & $-1.79 \pm 0.13$ & $-1.79 \pm 0.16$ & 0.72\\
1400 & 242.3979 & 54.0800 & $-1.35 \pm 0.58$ & $-1.33 \pm 0.93$ & 0.32\\
1452 & 242.3678 & 54.3307 & $-1.42 \pm 0.23$ & $-1.42 \pm 0.06$ & 0.76\\
1544 & 242.3142 & 54.2309 & $-1.31 \pm 0.07$ & $-1.31 \pm 0.01$ & 2.60\\
2150 & 241.9042 & 54.1795 & $-1.46 \pm 0.14$ & $-1.46 \pm 0.09$ & 0.49\\
2480 & 241.4090 & 54.6925 & $-1.37 \pm 0.09$ & $-1.38 \pm 0.06$ & 0.24\\

\hline
\end{tabular}
\end{table*}

\subsection{Spectral Classification}\label{sec:spec_class}

We have generated a radio colour plot for 697 sources located in the ELAIS N1 field that have matches in all four catalogues at frequencies: 146, 400, 612 and 1250 MHz and is displayed in Figure \ref{fig:radio_color}. To prevent any ambiguity, the plot does not show error bars. However, the median errors for $\alpha^{612}_{146}$ and $\alpha^{1250}_{400}$ are available as a reference, with values of 0.10 and 0.22, respectively. We have used the two-point spectral indices between 146\,MHz and 610\,MHz and between 400\,MHz and 1.25\,GHz to classify the sources into four spectral categories: 
\begin{enumerate}
    \item 
steep \& flat ($\alpha^{1250}_{400}\leq0$ \& $\alpha^{612}_{146} \leq0$), 
\item peaked ($\alpha^{1250}_{400}\leq0$ \& $\alpha^{612}_{146} > 0$), 
\item inverted ($\alpha^{1250}_{400}>0$ \& $\alpha^{612}_{146} > 0$), 
\item upturning ($\alpha^{1250}_{400}> 0$ \& $\alpha^{612}_{146} \leq0$)
\end{enumerate}

As evident, more sources lie in the steep and flat spectrum quadrant. Sources in the inverted quadrant are likely to be dominated by GPS or HFP sources whose spectral peaks lie at higher frequencies $\gtrsim 1$ GHz. The peaking quadrant consists of the sources whose turnover frequency lie within the frequency range 400-1400 MHz and is discussed in section \ref{MPS_sec}. The upturning quadrant comprises the composite sources that have a steep power-law spectrum at low frequencies and an inverted spectrum at high frequencies. This, thus, could be an indication of multiple epochs of AGN activity.

The majority of sources ($\sim 92$\,per\,cent) belong to the steep spectrum quadrant, and out of these, 29.5\,per\,cent are AGN, while the rest are SFGs. 
Out of the total 213 AGN shown in the figure, this steep quadrant covers almost 89\,per\,cent of the AGN population and is consistent with the properties of RL AGN. In the following subsections, we discuss the properties of USS and MPS sources in detail.

\begin{table}
    \centering
        \caption{Details of the potential USS source in the ELAIS N1 field.}
    \label{tab:USS_cand}
    \begin{tabular}{ccccc}
     \hline \hline \\
 ID & RA & DEC & $\alpha_{\rm 400\,MHz}^{\rm 1.2\,GHz}$ & $z$ \\ [1ex]
\hline 
2074 & 241.9562 & 54.8896 & $-1.31 \pm  0.27$  & 0.17\\
194 & 243.3113  & 54.2970  & $-1.64 \pm  0.10 $  & 2.87\\
344 & 243.1391 & 54.9420   & $-1.42 \pm  0.19$  & 0.08\\
2379 & 241.6505 & 54.5475 & $-1.34 \pm  0.06$  & 1.33\\
354 & 243.1264 & 54.8527  & $-2.17 \pm  0.07$  & 0.39\\
966 & 242.6523 & 54.6856  & $-1.43 \pm  0.04$  & 1.65\\
284 & 243.2046 & 55.0445  & $-1.64 \pm  0.07$  & 0.49\\
1316 & 242.4478 & 54.3250  & $-1.56 \pm  0.27$  & 0.34\\
479 & 243.0141 & 54.5015  & $-1.31 \pm  0.36$  & 0.45\\
1896 & 242.0792 & 54.8541 & $-1.70 \pm  0.27$  & 3.19\\

\hline
    \end{tabular}
\end{table}

\subsubsection{Ultra steep spectrum sources:}

High-redshift radio galaxies (HzRGs; $z>2$) are often observed in the early Universe and are typically located in the midst of protoclusters within regions of high density \citep{Roettgering_1994, Knopp_1997}. These overdense regions are conducive to galaxy formation and growth, making them an important target for studying the early stages of galaxy evolution and the physical processes driving them. HzRGs are thought to be the predecessors of the massive elliptical galaxies that are observed in the local Universe.
Ultra-steep spectrum sources (USS) tend to be favourable candidates for HzRGs with extremely steep spectral indices \citep{Blundell_1998,Miley_2008, Riseley_2016}. Different studies in the literature have used various selection limits on the spectral indices; for instance, \citet{Blundell_1998} used $\alpha_{\rm 151\,MHz}^{4.5\,{\rm GHz}}<-0.981$, \citet{DeBreuck_2004} : $\alpha_{\rm 843\,MHz}^{1.4\,{\rm GHz}}<-1.3$ and \cite{Singh_2014} : $\alpha_{\rm 325\,MHz}^{1.4\,{\rm GHz}}<-1.0$. The very steep alpha values observed in these sources are believed to be a result of radiation losses in the radio lobes from relativistic electrons \citep{Mahony_2016}, indicating that these sources are highly luminous at lower frequencies.

In this study, we employed the spectral limit of $\alpha<-1.3$ to identify USS sources from our sample in the ELAIS\,N1 field. The first method we used involved performing the SED fitting for the sources (see section \ref{sec:alpha}), while the second method involved using the radio colour plot to identify USS sources. We were able to identify 9 USS sources in our 400\,MHz catalogue with the SED fitting method. Table \ref{USS} presents the list of USS sources identified in this way. Out of these, one is identified as AGN while eight as SFGs from our previous analyses.
Meanwhile, using the radio-colour plot, we classified one source as a USS source, which was already identified through the SED fitting method. This source is shown as the open green square in Figure \ref{fig:radio_color}, and the spectra of all the nine sources are attached in the \ref{sec:USS_spectra} It is noted that excluding two sources, all the remaining sources in our study exhibit redshift values less than 1.

Besides, there are ten other sources for which $\alpha_{\rm 400\,MHz}^{1.25\,{\rm GHz}}<-1.3$ and are shown as open blue squares in Figure \ref{fig:radio_color}. Nevertheless, the alpha values obtained for these sources from the SED fitting are distributed in the redshift range 0.17--3.19. These could be potential USS sources in our uGMRT sample and the details of which are listed in Table \ref{tab:USS_cand}.

\subsubsection{Megahertz Peaked Spectrum Sources:}\label{MPS_sec}

We employed the radio colour-colour plot to identify sources with multiple power-law spectra in the ELAIS N1 field. Our criteria to select peaked spectrum sources were $\alpha_{\rm low} > 0.1, \rm{and}, \alpha_{\rm high} < -0.5$ \citep{Callingham_2017}, where $\alpha_{\rm low}$ and $\alpha_{\rm high}$ correspond to the spectral indices between 146\,MHz and 612\,MHz, and between 400\,MHz and 1.25\,GHz, respectively. The above limit is used to avoid contamination of flat spectrum sources and thus make the selection more reliable. Figure \ref{fig:radio_color} shows the red lines indicating the criteria used to identify the MPS sources. We successfully identified three sources meeting these criteria and have listed their details along with their respective redshifts in the first three rows of Table \ref{tab:MPS}. 

Moreover, in addition to the criteria based on the radio colour plot, we visually examined the power-law fits (Section \ref{sec:alpha}) of individual sources. This analysis led us to identify three more sources with spectral peaks at mega-hertz frequencies, which are not evident from the radio colour plot. The relevant information about these sources, along with their redshifts, is summarized in the bottom three rows of Table \ref{tab:MPS}. In order to perform spectral fitting on these sources separately, we utilized the generic model described in \cite{Callingham_2017}:
\begin{equation}
    S_\nu = \frac{S_{\rm p}}{(1-e^{-1})}\Big( 1 - e^{(\nu/\nu_{\rm p})^{\alpha_{\rm thin} - \alpha_{\rm thick }} }\Big) \Big(\nu/\nu_{\rm p}\Big)^{\alpha_{\rm thick}}
    \label{eq1}
\end{equation}
where $S_{\rm p}$ is the peak flux density at the peak frequency ($\nu_{p}$), $\alpha_{\rm thick}$ and $\alpha_{\rm thin}$ are the spectral indices in the optically thick and thin regimes of the spectrum.

\begin{figure}
  \centering
  \begin{tabular}{cc}
  \includegraphics[width=2.25in]{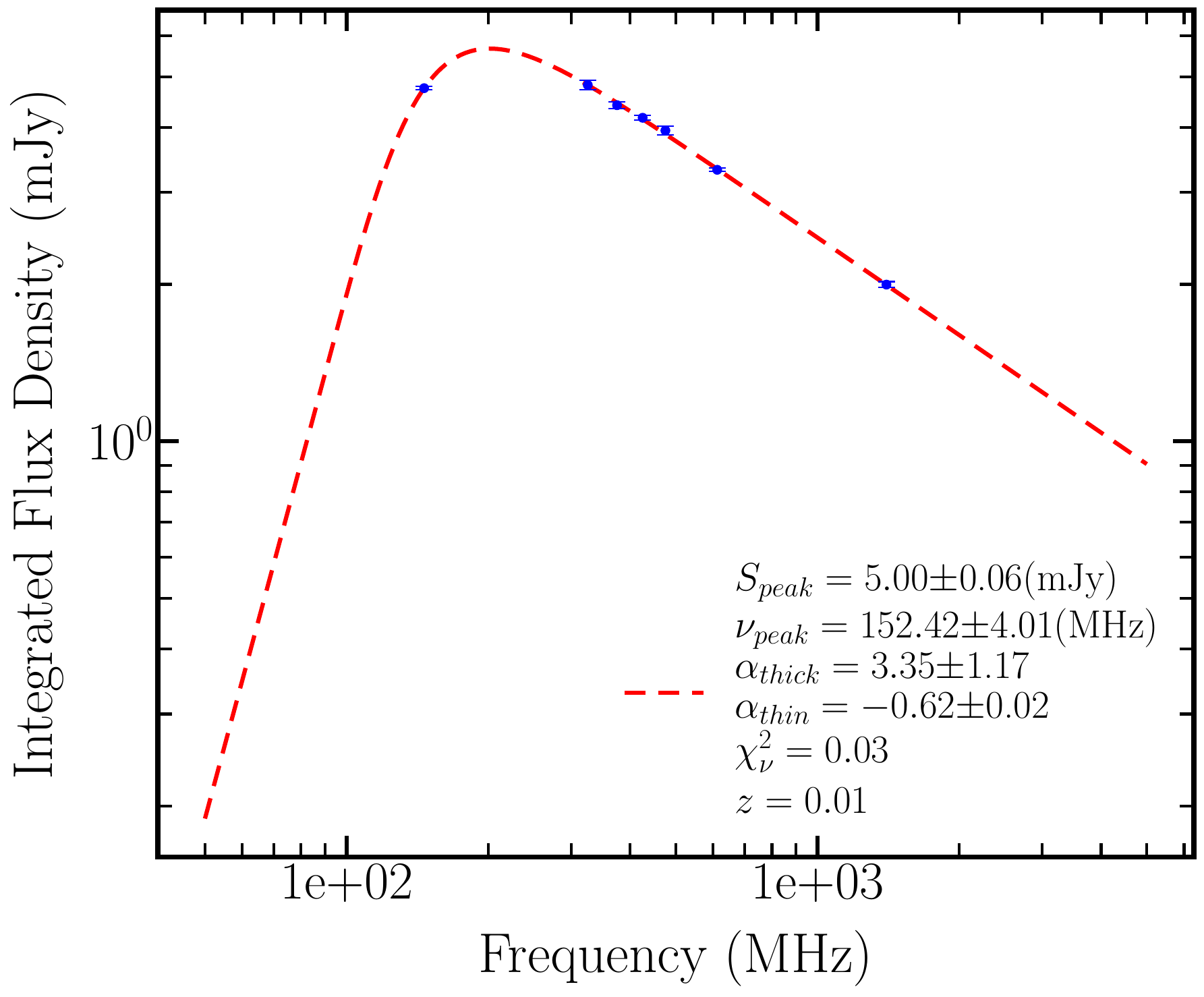} & \includegraphics[width = 2cm]{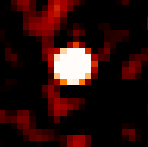} \\
   & \footnotesize{$S_{\rm{peak,400MHz}} = 4.005\,$mJy} \\
    \end{tabular}
    \begin{tabular}{cc}
  \includegraphics[width=2.25in]{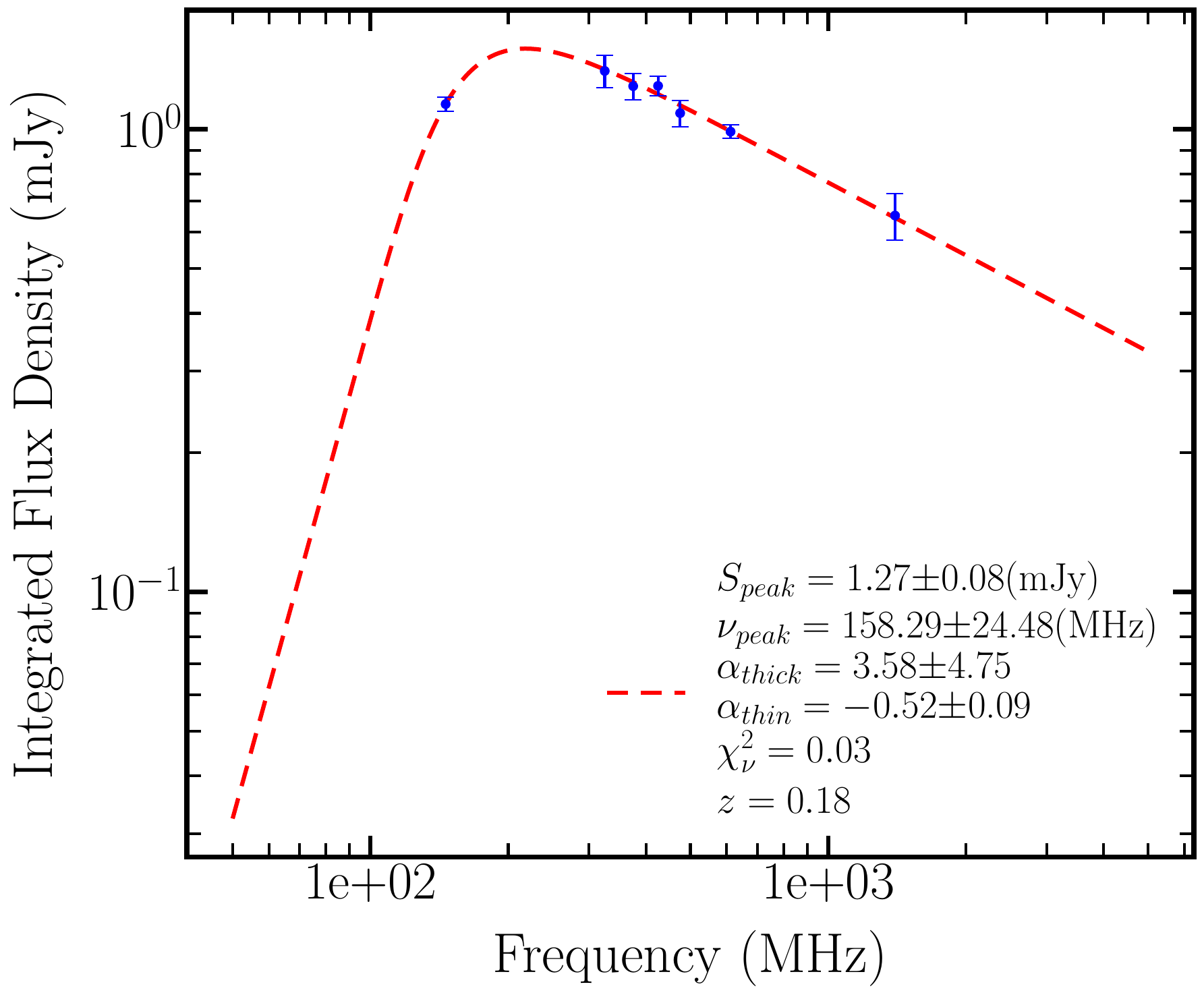} &
  \includegraphics[width = 2cm]{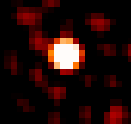} \\
   & \footnotesize{$S_{\rm{peak,400MHz}} = 1.2005\,$mJy} \\
    \end{tabular}
\begin{tabular}{cc}
  \includegraphics[width=2.25in]{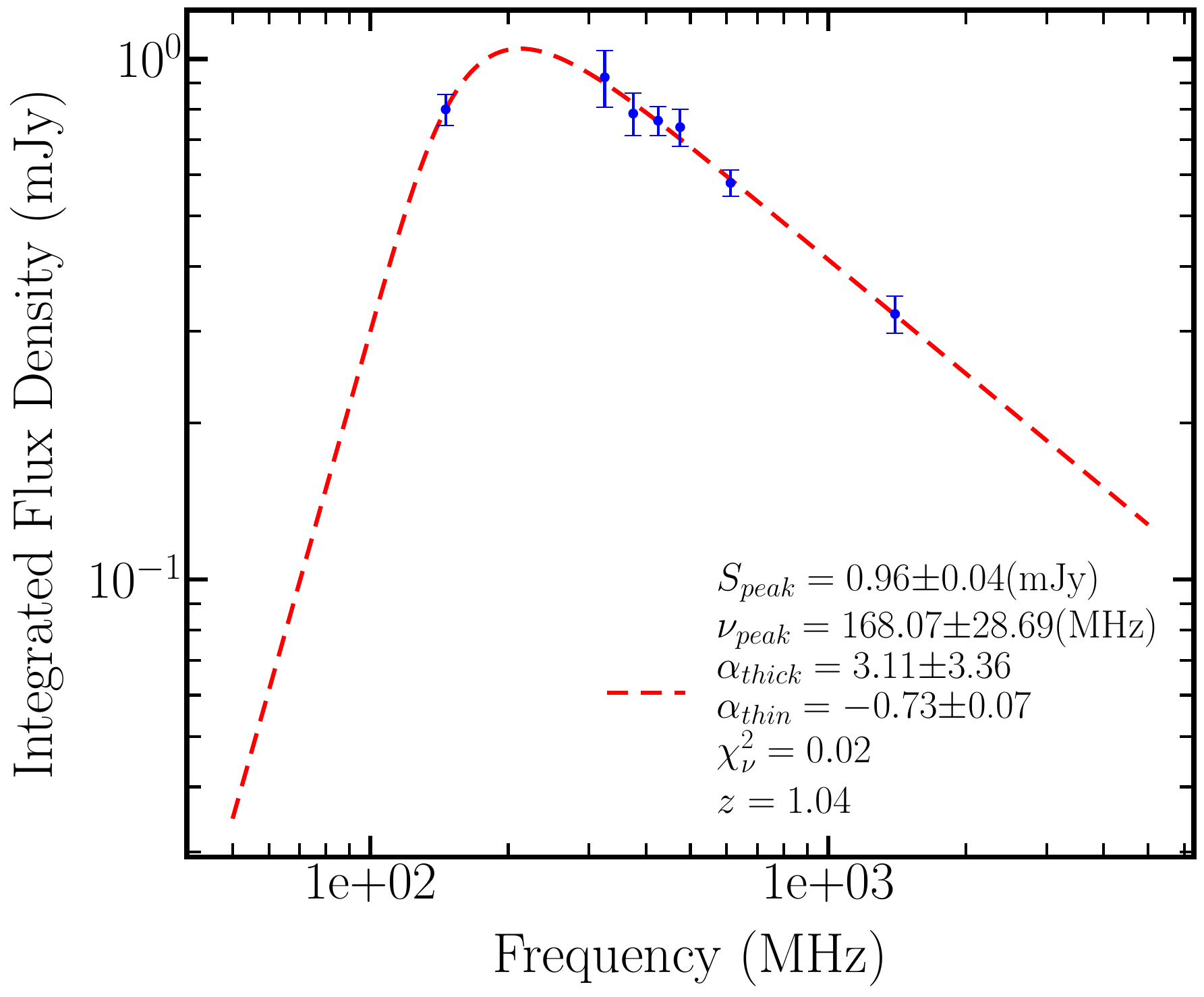} &
  \includegraphics[width = 2cm]{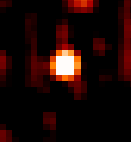} \\
   & \footnotesize{$S_{\rm{peak,400MHz}} = 0.7968\,$mJy} \\
    \end{tabular}

  \caption{Spectral energy distribution (left) for three MPS sources (source IDs 1155, 595 and 1194 in Table \ref{tab:MPS}) shown at 400\,MHz on the right. The $S_{\rm{peak,400MHz}}$ represents the peak flux values for these images at 400\,MHz.}
  \label{fig:mps_spectra}
\end{figure}

The top-panel of Figure \ref{fig:mps_spectra} displays the observed spectrum of the source with ID 1155 at a redshift of 0.01. As depicted from the image of Figure \ref{fig:mps_spectra}, it is a compact source at 400\,MHz. The measured peak frequency of this source is $152.4\pm4.01$\,MHz, while the values of the spectral indices $\alpha_{\rm thick}$ and $\alpha_{\rm thin}$ are $3.35\pm1.17$ and $-0.62\pm0.02$, respectively.

Also, the middle and bottom panels of Figure \ref{fig:mps_spectra} present the fitted spectra for sources with ID 595 and 1194 and having their peak frequencies around $158.29\pm24.48$\,MHz and $168.07\pm28.69$\,MHz, respectively. These sources are also compact and  detected at redshifts 0.18 and 1.04.

\begin{table}
    \centering
    \caption{Details of the MPS sources in the ELAIS N1 field. The top three rows in the table represent the MPS sources that were identified using the radio colour diagram. On the other hand, the bottom three sources correspond to the MPS sources that were selected based on visual inspection from the SED fits. }
    \begin{tabular}{cccc}
    \hline \hline
    ID & RA & DEC & $z$ \\
     & (deg) & (deg) &   \\ [1ex]
    \hline 
    1022 & 242.62222 & 54.92758 & 0.267 \\
    1925 & 242.05520 & 55.00188 & 1.74  \\
    2162 & 241.89594 & 54.55692 & 3.38  \\
    \hline
    1155 & 242.55263 & 54.59195 & 0.013  \\
    595 & 242.91821 & 54.24502 & 0.18  \\
    1194 & 242.52626 & 54.30840 & 1.04  \\

    \hline
      \end{tabular}
    \label{tab:MPS}
\end{table}

The MPS sources identified in this study were found to span a redshift range of $0.01<z<3.38$, with only one source having spectroscopic redshift information. This range suggests that these sources could potentially be young AGN at higher redshifts. 
It has been previously reported that MPS sources can encompass a combination of nearby CSS and GPS, and HFP sources at high redshifts \citep{Coppejans_2015}. This is also likely the case for the sources included in our study. 
The parameters derived for the three MPS sources using eq. \ref{eq1} are similar to those presented in \cite{Keim2019}, where the authors suggest FFA as the most likely absorption mechanism based on the source's peak frequencies, linear sizes, and magnetic field. 
Therefore, it is possible that the bottom three sources in Table \ref{tab:MPS} also exhibit FFA as the absorption mechanism. However, more data are required at the low frequencies to get robust SED fitting parameters and therefore to comprehend the physical mechanism responsible for the observed peaked spectra. On the other hand, it is also important to note that for the SSA mechanism to occur, the spectral slope cannot cross the threshold of $+2.5$ below the turnover and therefore, FFA has been the widely suggested mechanism for any such scenarios \citep[see for e.g.][]{Bicknell_1997, Callingham_2015}.

\subsection{Radio spectral index vs. redshift}

It is believed that there lies a strong anti-correlation between the spectral index and redshift of the radio sources \citep{Miley_2008,Athreya_1999}. The study of z$\sim \alpha$ correlation has been used to search for high-redshift radio galaxy (HzRG) candidates in large area radio surveys (see \cite{Rottgering_1997}, for instance). The plausible explanations of z$ \sim \alpha$ correlation i.e., the steepness of spectral index with redshift, are as follows: (a) K-correction: 
In the case of high redshift sources, the spectrum experiences a shift towards lower frequencies, resulting in the inclusion of the steep portion within the observed spectrum.
(b) Indirect manifestation of luminosity, L$\sim \alpha$ effect \citep{Chambers_1990}. (c) Density-dependent effect: The ambient density increases at higher redshifts; thus, jets from radio sources have to move against the surrounding denser medium, which in turn gives rise to strong synchrotron losses at higher frequencies.

The studies by \cite{Morabito_2018} in radio galaxies suggested that the observed $z\sim \alpha$ relation could be explained by the possibility of combining the selection effects and inverse Compton losses at high redshifts. Whereas, \cite{Saxena_2019} studied a sample of 32 USS sources and found no strong anticorrelation between $z$ and $\alpha$ among their sample.

\begin{figure}
    \centering
    \includegraphics[width=\columnwidth]{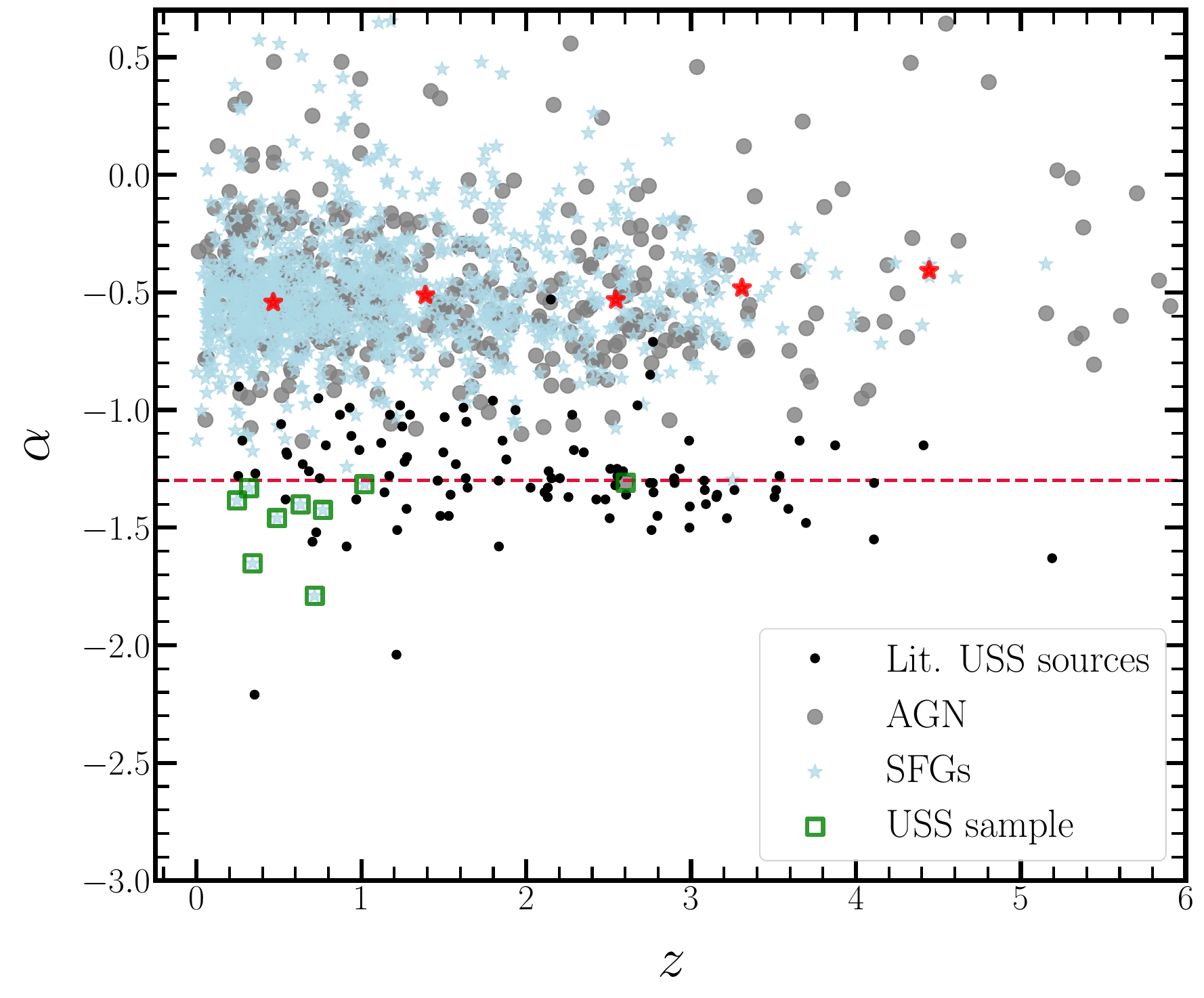}
    \caption{Radio Spectral index as a function of redshift for the SFGs, AGN and USS sources in the ELAIS N1. The red dashed line represents $\alpha=-1.3$, the criterion adopted to identify USS sources.}
    \label{fig:alpha_z}
\end{figure}

We study the observed correlation between radio spectral index and redshift for the sources in the ELAIS N1 field. Figure \ref{fig:alpha_z} represents the variation of radio spectral index with redshift for AGN (grey circles), SFGs (light-blue stars) and USS sources (open green squares). Here, the spectral indices used are the ones derived using the SED fits. The Pearson correlation coefficient (r) for the SFGs, AGN and USS sources in Figure \ref{fig:alpha_z} are 0.08, -0.15 and 0.28, respectively. We don't find any strong anti-correlation for the radio sources in the ELAIS N1 field. The limited number of USS samples we obtained from the field indicate no observable changes in spectral indices as redshift increases.

Besides, the SFGs sample shows an almost negligible variation of spectral indices with redshift. This is consistent with the analyses of \cite{Magnelli_2015}, \cite{Ivison_2010} and \cite{Calistro_2017}.
The red star symbols depicted in Figure \ref{fig:alpha_z} correspond to the median values of SFGs across five redshift bins. 
This non-evolution of spectral values with redshift suggests that the radio spectrum is independent of the properties of the galaxies in the large-scale context where redshift evolution plays a major role. This further implies that the local properties of the galaxies like magnetic fields, surrounding interstellar medium (ISM) and cosmic ray electrons (CREs) from the supernova contribute to the nature of the radio SED.

\vspace{-1em}
\section{Discussion \& Summary}

In this study, we focused on the ELAIS N1 field of the deep extra-galactic sky and analyzed it at 1250\,MHz using the uGMRT. We reached an RMS noise of $\sim 12\mu\rm{Jy\,beam^{-1}}$ and catalogued 1086 sources at this frequency. 
Given that the ELAIS N1 field has been widely explored at various frequencies, we cross-matched our uGMRT sample at 400\,MHz with other catalogues to investigate their various spectral properties. 

From our analyses of the radio SED fits, we determined the median spectral index value for our uGMRT sample to be $\alpha =-0.57\pm0.14$ which implies the flattening of radio spectral values at low frequencies. Whereas, the median value of the two-point spectral indices measured between the frequency 400\,MHz and 1250\,MHz is $\alpha_{400}^{1250} = -0.66\pm0.17$. After cross-matching sources in the uGMRT 400 MHz with other radio catalogues at different frequencies, we found that the majority of the sample belongs to the steep spectrum quadrant with less number of sources in other quadrants, as depicted in Figure \ref{fig:radio_color}. This Figure presents the radio colour diagram for SFGs and AGN, the classification of which is obtained from \cite{Akriti_2022} at 400\,MHz. 
Based on the spectral indices measured by employing the SED fits, we identify nine USS sources in the redshift range of 0.24--2.60.

Furthermore, based on the radio colour diagram, we classify three sources as MPS sources. Moreover, three additional sources were visually selected due to their exhibited peaked spectra at MHz frequencies. 
A general SED fit was performed to determine their peaked frequencies that lie around $152, 158$ and 168 \,MHz. 
The measured spectral slope below the turnover for all three sources is found to be $\alpha_{\rm thick}>3.1$. This value indicates that the FFA mechanism is the plausible explanation for the observed spectra in these cases, but for a detailed analysis, more data are required, especially below the spectral turnover.

Finally, in our study, we present the analysis of spectral indices (obtained through SED fitting) and their correlation with redshifts for three distinct source categories: SFGs, AGN and USS sources. We observe a lack of strong anti-correlation among the radio sources in the ELAIS N1 field. Especially for SFGs, this may mean that the nature of the radio SED is mostly dependent on the local parameters within the galaxies, like magnetic fields, properties of the surrounding ISM etc. and is independent of the properties in the large-scale context for which redshift evolution becomes crucial. 


\appendix

\begin{figure*}
    \centering
    \includegraphics[height = 7cm]{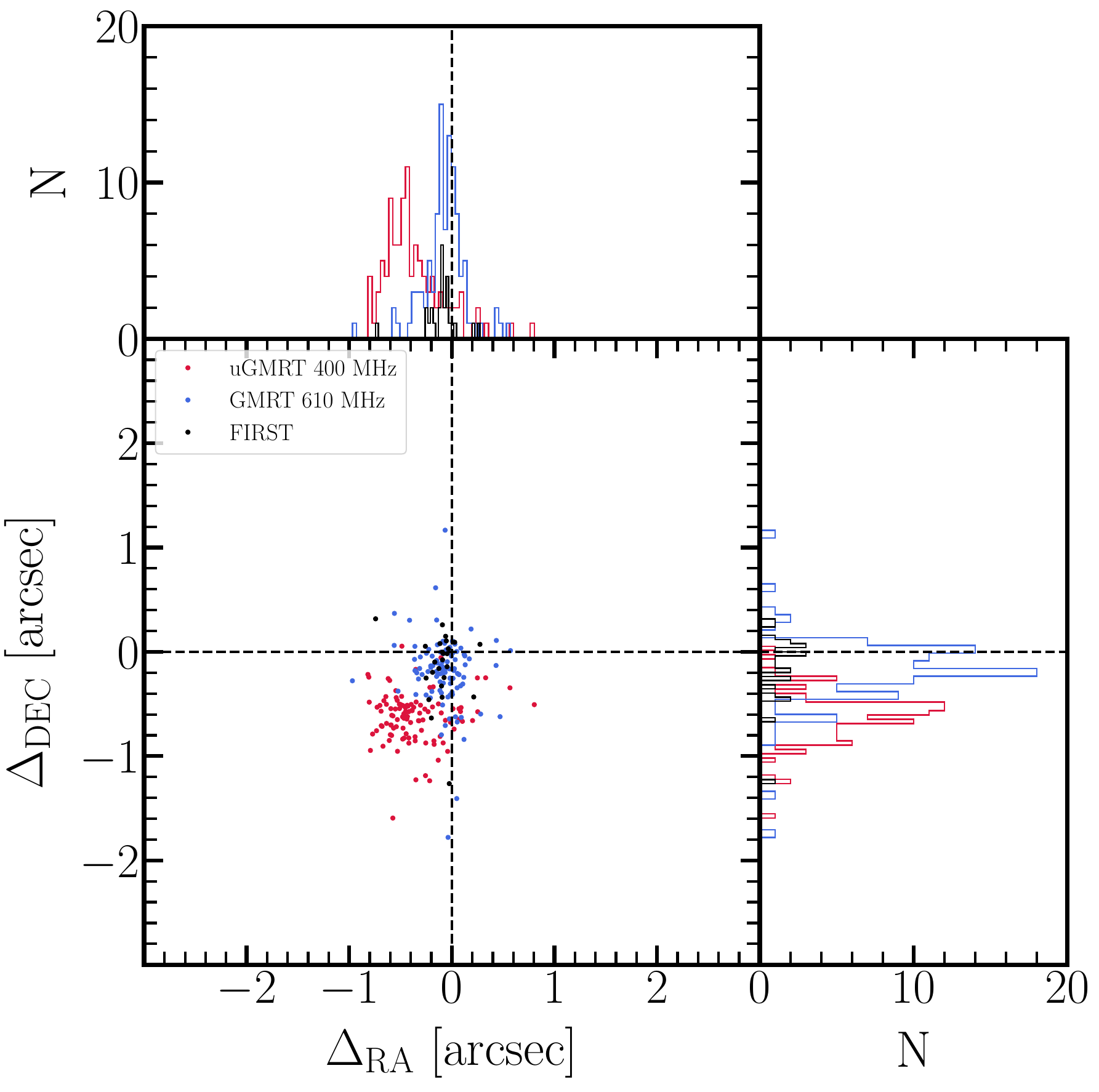}
    \includegraphics[height = 7cm]{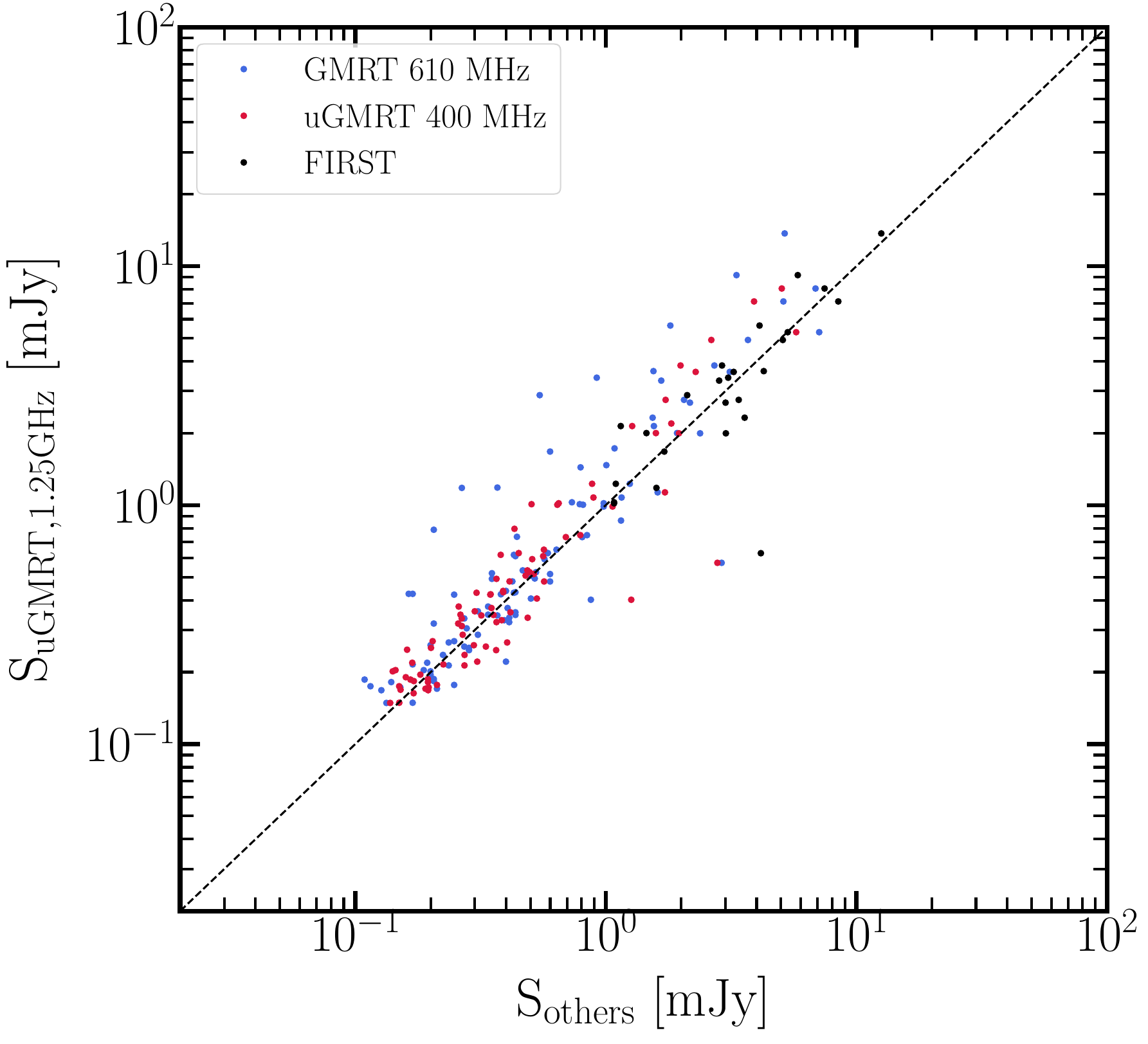}

    \caption{\textit{Left:} The positional offset for the uGMRT sample at 1.25\,GHz from FIRST (black), GMRT 610\,MHz (blue) and uGMRT 400\,MHz (red) catalogues. \textit{Right:} Variation of integrated flux densities at 1.25\,GHz with other radio catalogues scaled to this frequency. The colour schemes are the same as in the left panel.}
    \label{fig:comp}
\end{figure*}

\section{Positional \& Flux Accuracies}\label{sec:flux_position}

Here, we compare the uGMRT 1.25\,GHz catalogue to the other radio catalogues in the literature. We have used the 1.4\,GHz Faint Images of
the Radio Sky at Twenty centimetres (FIRST) survey \citep{White_1997}, uGMRT catalogue at 400\,MHz from \citep[see Section \ref{sec:survey},][for details]{Arnab2019} and the GMRT catalogue at 610\,MHz by \cite{Ishwara_2020} with the resolution of $6''$. We used a search radius of $2''.0$ to identify a cross-match in other catalogues. For the positional and flux accuracy analysis, we have applied a sample selection criteria of sources following \cite{Williams_2016}: high signal-to-noise ratio ($>10$) sources, compact sources with size less than the resolution of the catalogue and isolated sources for which the minimum distance between two sources are greater than twice of the resolution.

\subsection{Positional accuracy}
The positional offsets in right ascension (RA) and declination (DEC) for the uGMRT sample at 1.25\,GHz are measured as:
\begin{align}
\begin{split}
    \delta_{\rm RA} &=  \rm{RA}_{\rm uGMRT}  - \rm{RA}_{\rm FIRST} \\
    \delta_{\rm DEC} &=  \rm{DEC}_{\rm uGMRT} - \rm{DEC}_{\rm FIRST}
\end{split}
\end{align}
The FIRST catalogue has positional accuracy better than $1''$ with a resolution of $\sim5''$. We measure the median values in the deviation of RA and DEC using the FIRST catalogue as $-0.092''$ and $-0.076''$, respectively. Fig. \ref{fig:comp} (left) presents the offsets in RA and DEC for the uGMRT source catalogue compared to the other catalogues, along with their histograms. The median offsets in RA, DEC, as measured from the GMRT 610\,MHz and the uGMRT 400\,MHz catalogues, are -0.07, -0.16 and -0.43, 0.59, respectively. It should be noted that the resolution of our catalogue $\sim2''$ is better than the resolution of other catalogues $\sim 5''-6''$, and the median offset with the FIRST catalogue is less than $0.1''$. Hence we do not apply any corrections in the source positions in our uGMRT catalogue.

\subsection{Flux accuracy}

Our uGMRT 1.25\,GHz catalogue was generated using \cite{Perley_2017} flux scales. Each catalogue will have different flux scales, therefore, we have made sure to convert them to the flux scales used in our work. We measure the ratio of the integrated flux density at 1.25\,GHz with the other catalogues also scaled to 1.25\,GHz using a constant spectral index value of -0.7. This ratio is defined as $S_{\rm{1.25\,GHz}}/ S_{\rm{other}}$. In Fig. \ref{fig:comp} (right), we show the comparison of the $S_{\rm{1.25\,GHz}}$ with $S_{\rm{other}}$ and no significant deviation is observed from the $S_{\rm{1.25\,GHz}}/S_{\rm{other}}=1$ line (black dashed line). The median $S_{\rm{1.25\,GHz}}/S_{\rm{other}}$ ratio as derived using the FIRST, uGMRT 400\,MHz and GMRT 610\,MHz catalogues are $0.99^{0.19}_{-0.37}$, $1.11_{-0.51}^{0.25}$ and $1.10_{-0.9}^{0.32}$, respectively. The errors quoted here are from the 16th and 84th percentiles. The median of the ratio is approximately 1 for these cases and therefore we do not suggest any correction for systematic offsets.

\section{Spectra of USS Sample} \label{sec:USS_spectra}

\begin{figure*}
    \centering
    \begin{tabular}{ccc}
      \includegraphics[scale = 0.18]{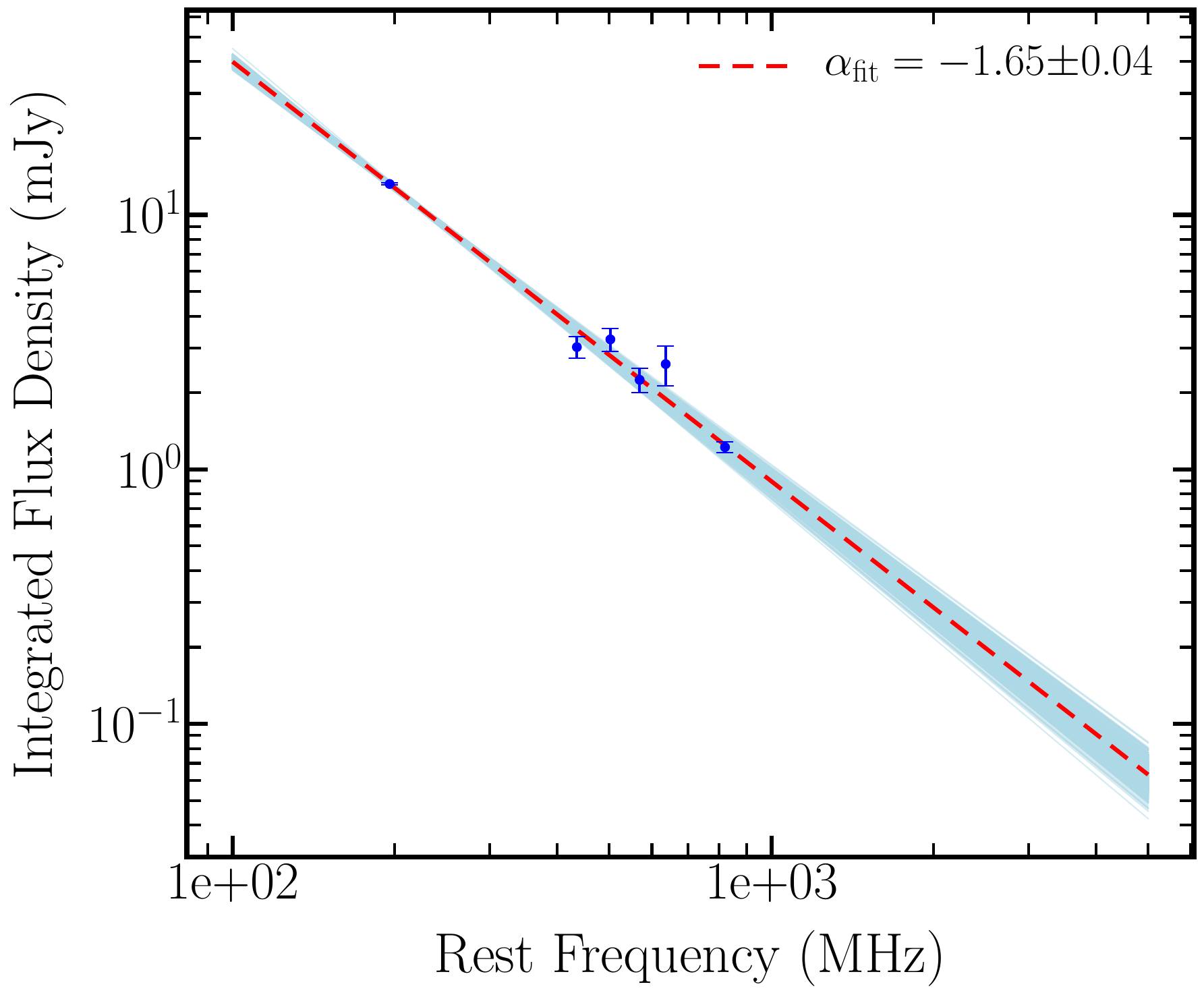} &
    \includegraphics[scale = 0.18]{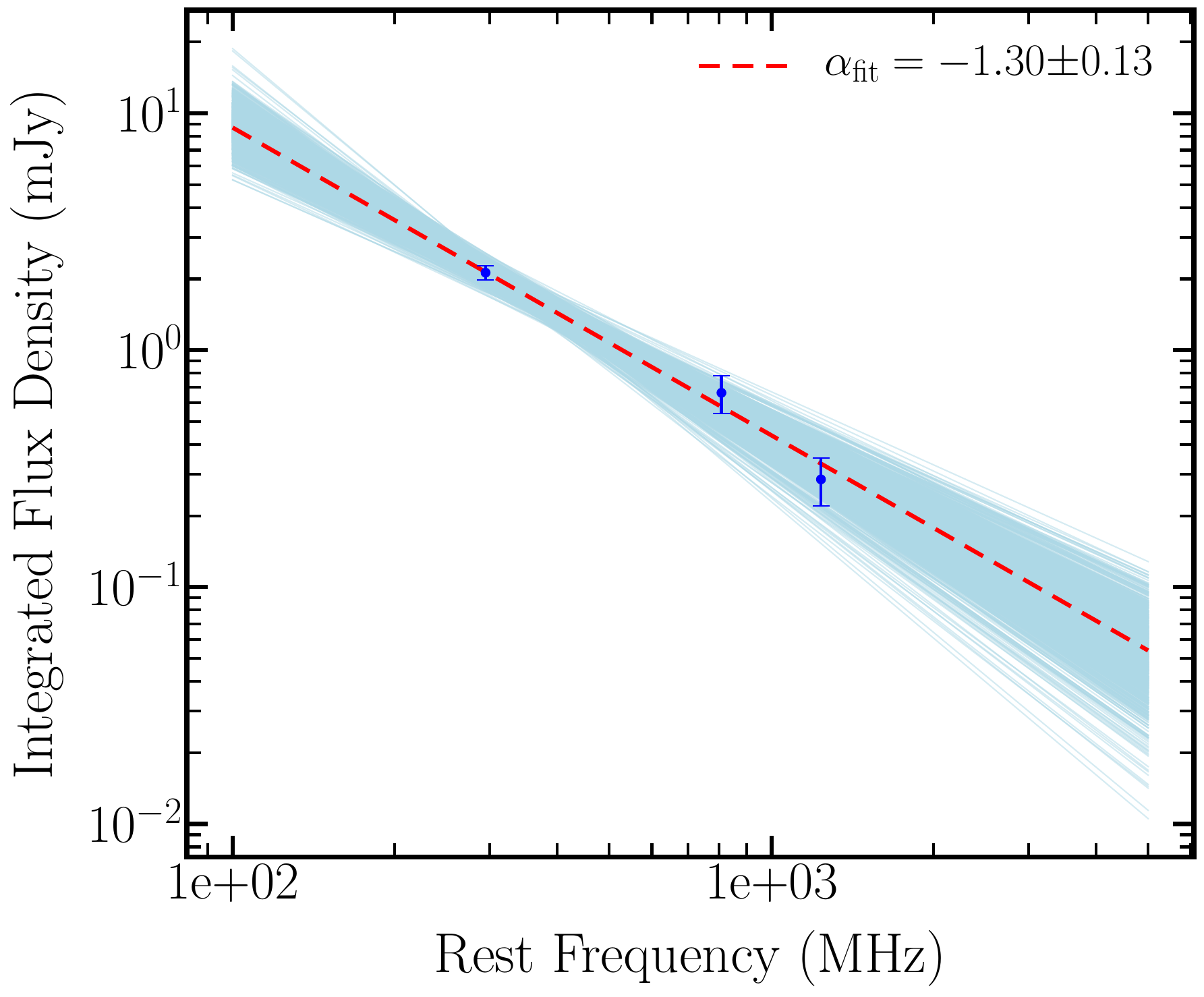} &
    \includegraphics[scale = 0.18]{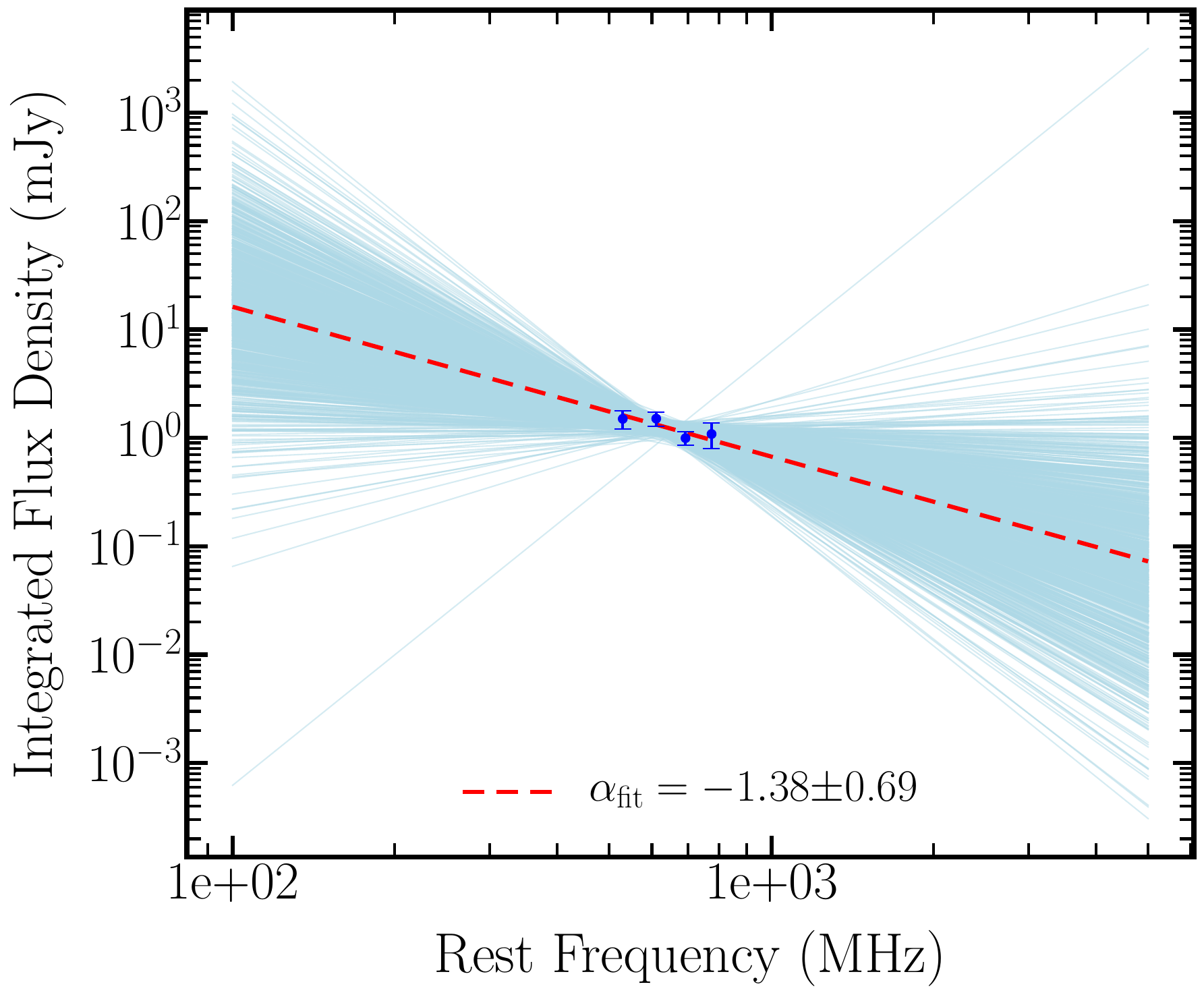} \\
    Source ID: 50 & Source ID: 1180 & Source ID: 1335\\
   & & \\
    \includegraphics[scale = 0.18] {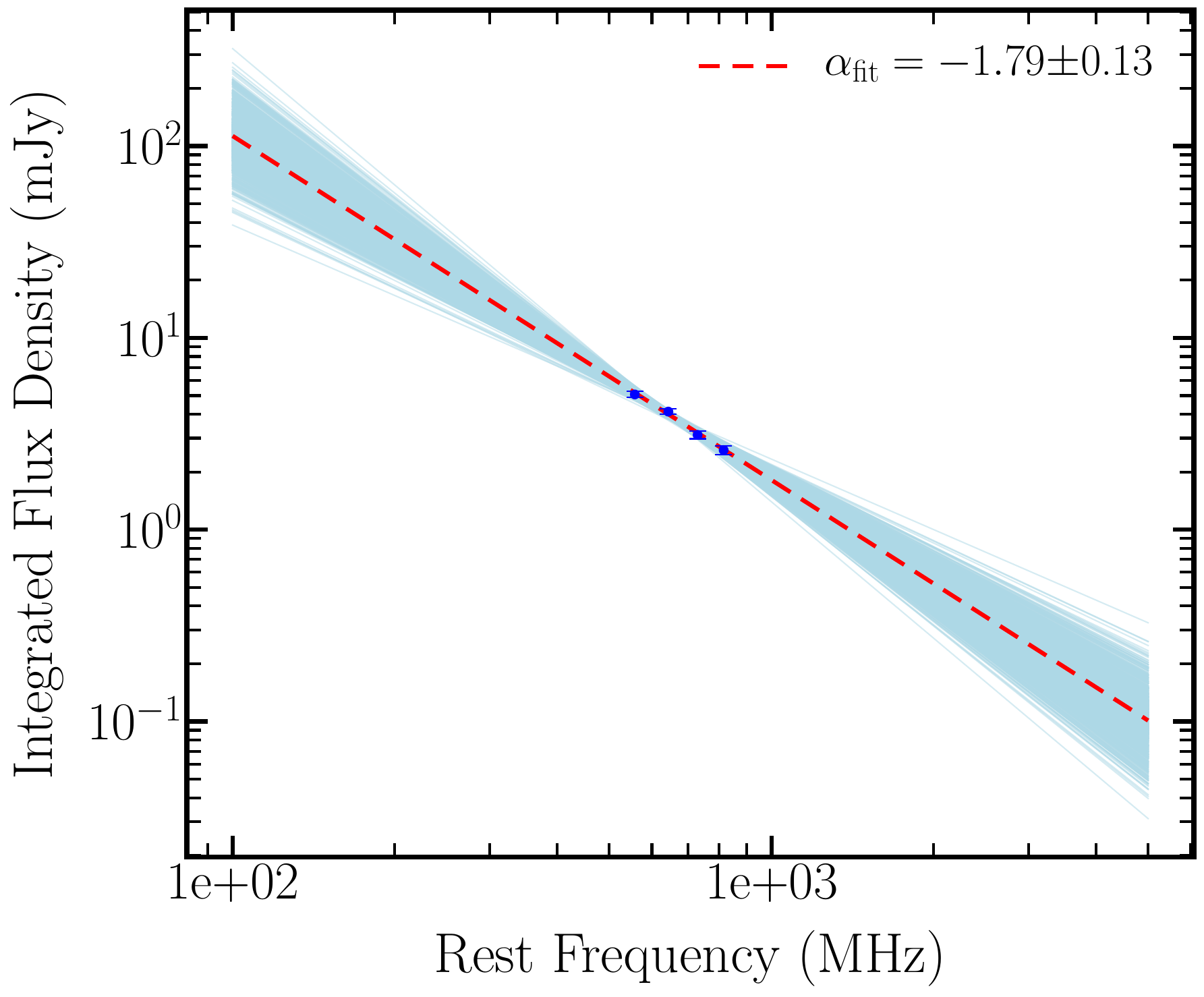} &
    \includegraphics[scale = 0.18]{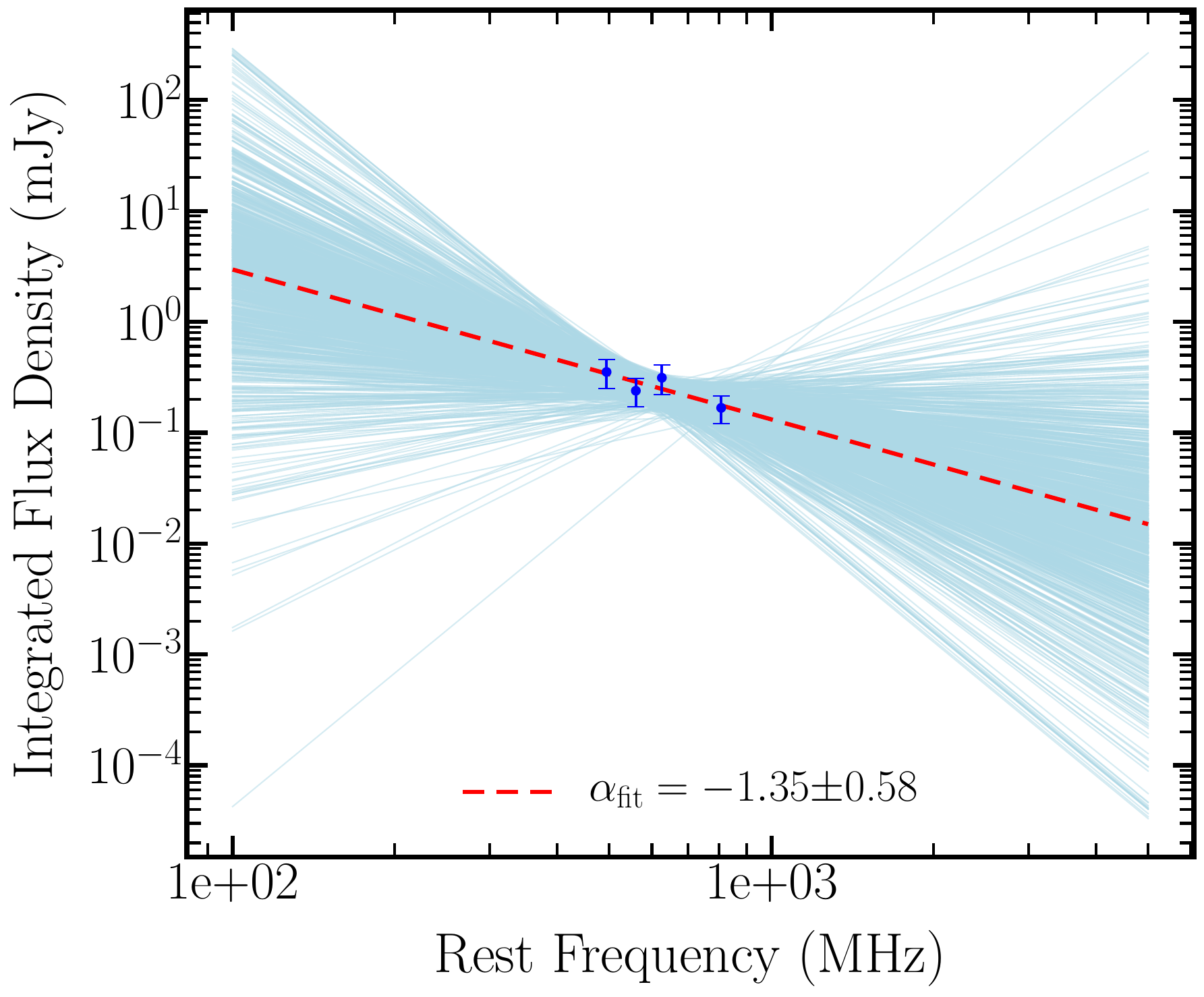} &
    \includegraphics[scale = 0.18]{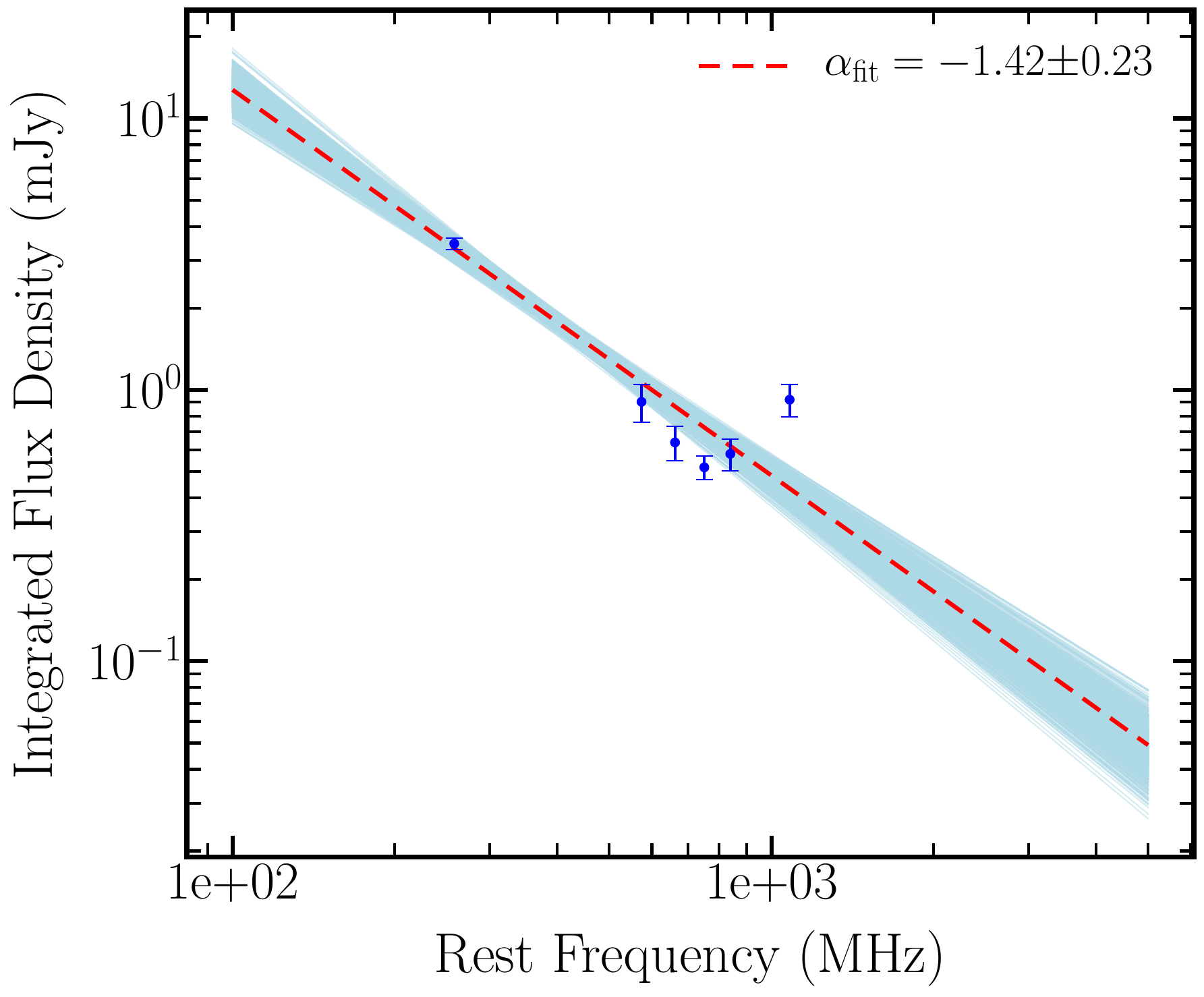} \\
    Source ID: 1344 & Source ID: 1400 & Source ID: 1452\\
       & & \\

    \includegraphics[scale = 0.18]{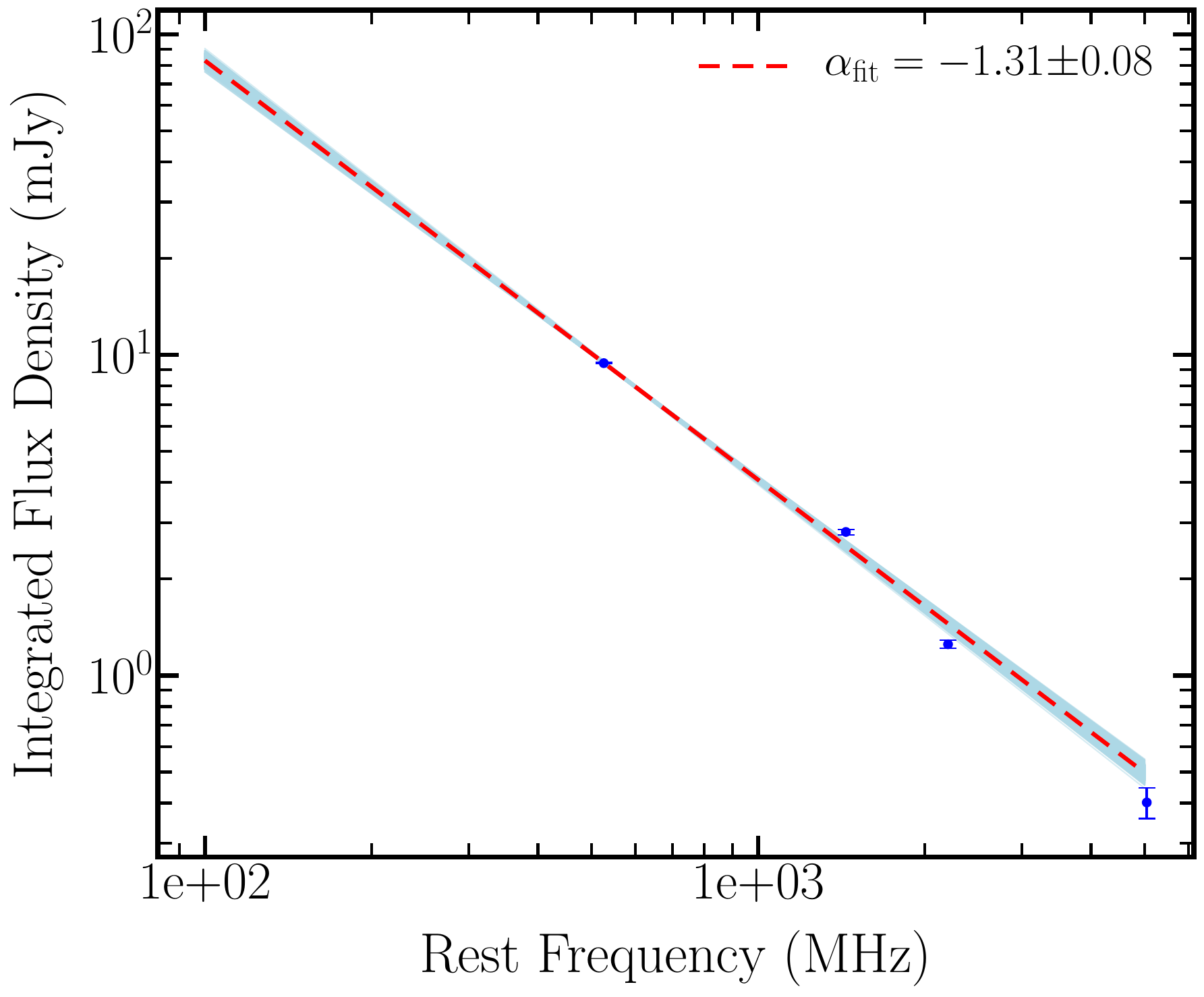} &
    \includegraphics[scale = 0.18]{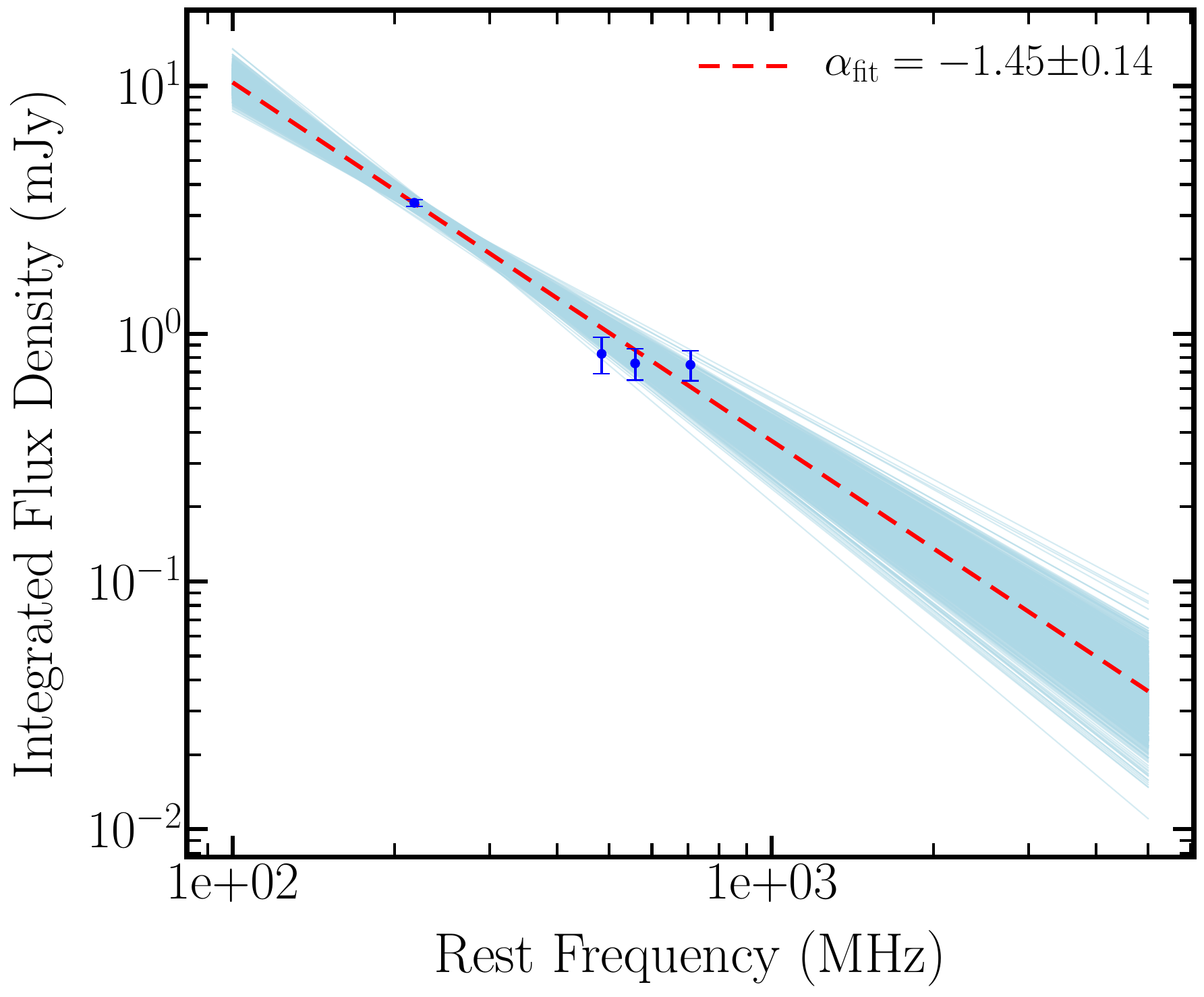} &
    \includegraphics[scale = 0.18]{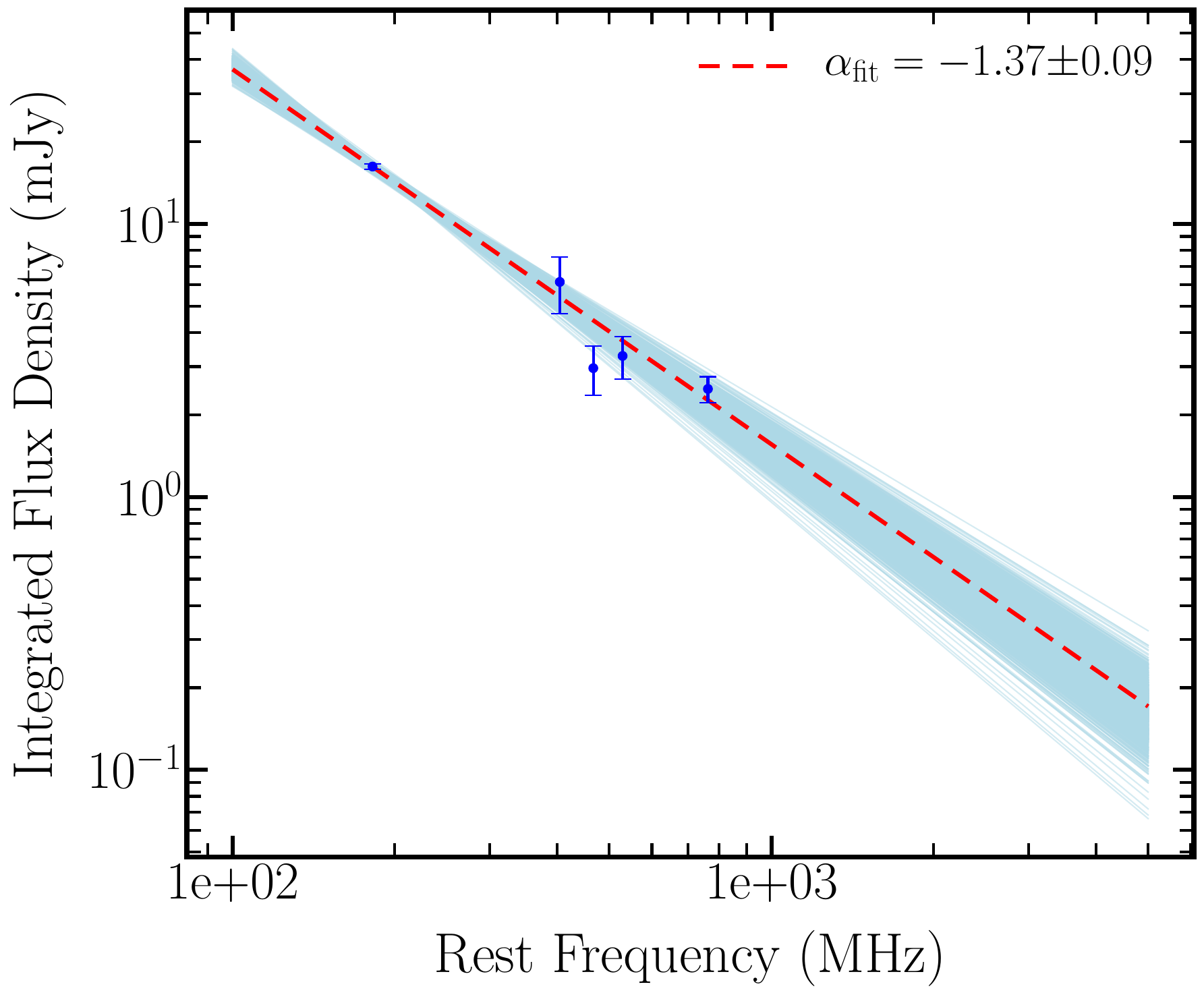} \\
    Source ID: 1544 & Source ID: 2150 & Source ID: 2480\\
    \end{tabular}

    \caption{The spectra for the USS sources in the ELAIS N1 field. The source IDs mentioned here are from the 400\,MHz uGMRT catalogue. }
    \label{fig:USS_spectra}
\end{figure*}

The spectra of the USS sources are shown in Figure \ref{fig:USS_spectra}. 

\section*{Acknowledgements}

We thank the anonymous referee for their comments
on the manuscript.
We further would like to thank Arnab Chakraborty for his helpful suggestions.
AS would like to thank DST for INSPIRE fellowship. 
We thank the staff of GMRT for making this observation possible. GMRT is run by National Centre for Radio Astrophysics of the Tata Institute of Fundamental Research. 

\vspace{-1em}
\bibliography{ref}





\end{document}